%% file: DRJCA.tex
\documentclass[11pt,a4paper]{article}
\usepackage{jheppub}
\usepackage{amsmath,amssymb}
\usepackage{graphicx}
\usepackage{color}
\usepackage[table]{xcolor}
\usepackage{slashed}
\usepackage{url}
\usepackage{enumitem}
\usepackage{multirow}
\usepackage{array}
\usepackage{tabularx}
\usepackage{float}
\usepackage{subcaption}
\usepackage{afterpage}

\newcommand{\dsum}{\displaystyle\sum}
\newcommand{\sumab}{\dsum_{a<b}}

\newcommand{\ith}{i^\text{th}}
\newcommand{\jth}{j^\text{th}}
\newcommand{\kt}{k_t}

\newcommand{\dij}{d_{ij}}
\newcommand{\diB}{d_{iB}}

\newcommand{\pt}{p_T}
\newcommand{\pti}{p_{T_i}}
\newcommand{\ptj}{p_{T_j}}
\newcommand{\pta}{p_{T_a}}
\newcommand{\ptb}{p_{T_b}}

\newcommand{\dR}{\Delta R}
\newcommand{\dRij}{\Delta R_{ij}}
\newcommand{\dRab}{\Delta R_{ab}}

\newcommand{\sigmaR}{\sigma}
\newcommand{\sigmaRi}{\sigma_i}
\newcommand{\sigmaRj}{\sigma_j}

\newcommand{\Rd}{R_d}
\newcommand{\Rdi}{R_{d_i}}

\newcommand{\Tops}{{\it tops}}
\newcommand{\Top}{{\it top}}
\newcommand{\topAone}{{\it top}~(A1)}
\newcommand{\topAtwo}{{\it top}~(A2)}

\newcommand{\V}{{\it V}}
\newcommand{\VBone}{{\it V}~(B1)}
\newcommand{\VBtwo}{{\it V}~(B2)}

\newcommand{\jets}{{\it jets}}
\newcommand{\jet}{{\it jet}}
\newcommand{\jetAone}{{\it jet}~(A1)}
\newcommand{\jetAtwo}{{\it jet}~(A2)}
\newcommand{\jetBone}{{\it jet}~(B1)}
\newcommand{\jetBtwo}{{\it jet}~(B2)}

\newcommand{\softjet}{{\it soft jet}}
\newcommand{\soft}{{\it soft}}
\newcommand{\softjets}{{\it soft jets}}

\newcommand{\MGparton}{MG5 parton}
\newcommand{\MGpartonlevel}{MG5 parton-level\ }

\newcommand{\W}{{\it W}}

\newcommand{\Rnot}{R_0}
\newcommand{\Acc}{${\cal A}\ $}
\newcommand{\suma}{\dsum_a}
\newcommand{\pnew}{p_q}
\newcommand{\ptnew}{p_{T_q}}
\newcommand{\pr}{p_r}
\newcommand{\ptr}{p_{T_r}}
\newcommand{\ps}{p_s}
\newcommand{\pts}{p_{T_s}}
\newcommand{\dRanew}{\Delta R_{aq}}
\newcommand{\dRar}{\Delta R_{ar}}
\newcommand{\dRas}{\Delta R_{as}}
\newcommand{\dRrs}{\Delta R_{rs}}
\begin{document}

%===============================================================================
\title{Dynamic Radius Jet Clustering Algorithm}

\author[a]{Biswarup Mukhopadhyaya}
\emailAdd{biswarup@iiserkol.ac.in}

\author[a]{\!\!, Tousik Samui}
\emailAdd{tousiksamui@gmail.com}

\author[a]{\!\!, and Ritesh K. Singh}
\emailAdd{ritesh.singh@iiserkol.ac.in}
\affiliation[a]{Department of Physical Sciences, Indian Institute of Science
Education and Research Kolkata, Mohanpur, 741246, India.}

%\date{\today}
%===============================================================================

%===============================================================================
\abstract{
The study of standard QCD jets produced along with fat jets, which may appear as
a result of the decay of a heavy particle, has become an essential part of
collider studies. Current jet clustering algorithms, which use a fixed radius
parameter for the formation of jets from the hadrons of an event, may be
inadequate to capture the differing radius features. In this work, we develop an
alternative jet clustering algorithm that allows the radius to vary dynamically
based on local kinematics and distribution in the $\eta$-$\phi$ plane inside
each evolving jet. We present the usefulness of this dynamic radius clustering
algorithm through two Standard Model processes, and thereafter illustrate it for
a scenario beyond the Standard Model at the 13~TeV LHC.}
%===============================================================================

%%%%%%%%%%%%%%%%%%%%%%%%%%%%%%%%%%%%%%%%%%%%%%%%%%%%%%%%%%%%%%%%%%%%%%%%%%%%%%%%
\maketitle
%%%%%%%%%%%%%%%%%%%%%%%%%%%%%%%%%%%%%%%%%%%%%%%%%%%%%%%%%%%%%%%%%%%%%%%%%%%%%%%%
\vspace{-10mm}
\section{Introduction}\label{sec:intro}
\vspace{-3mm}
The physics extraction capacity of any high-energy collider depends crucially on
the handling of coloured particles in various final states. These are produced
as partons via either short-distance interactions of quantum chromodynamics
(QCD) or electroweak processes\,\cite{Campbell:2006wx,Ellis:1996mzs}. The
partons, however, hadronize through long-distance QCD effects which are not
calculable {\em ab initio}. One rather uses semi-empirical methods to predict
the probability that energetic partons will fragment into more low-energy
partons and ultimately form colour-neutral hadrons which are observable in the
detector. Groups of closely spaced hadrons with varied degrees of collimation
form `jets' whose identification, isolation, and merger are predicted once more
with the help of semi-empirical (and by no means uniquely decided) algorithms
called jet clustering
algorithms\,\cite{Sterman:1977wj,Huth:1990mi,Blazey:2000qt,Ellis:2007ib,Salam:2010nqg}.
The aim always remains to define jets with such algorithms which most accurately
elicit the short-distance physics underlying the events that are studied. They
thus constitute some of our most important tools in the analysis of phenomena at
colliders.

In the context of the Large Hadron Collider (LHC), a widely used class of jet
criteria is based on so-called $\kt$-type sequential recombination jet
algorithms\,\cite{Catani:1991hj,Catani:1993hr,Ellis:1993tq,Dokshitzer:1997in,Wobisch:1998wt,Cacciari:2008gp,Salam:2010nqg}.
These algorithms (briefly discussed in the next section) typically try to merge
`neighbouring' hadrons to identify the group as a jet. The neighbourhood of a
hadron is defined by a single radius parameter $\Rnot$ in the $\eta$-$\phi$
plane of the detector, which is used to quantify the radius (or size) of a jet.
This is because the hadrons within $\Rnot$ are merged to form a jet while the
hadrons outside $\Rnot$ are not included in that jet. The choices for the value
of $\Rnot$ in these algorithms depend on the physics searches one is carrying
out. At the 13~TeV LHC, the typical choices for $\Rnot$ are $0.4$ or $0.8$ for a
`narrow' or a `fat' jet, respectively. There are, in addition, jet isolation
criteria depending on whether one is trying to separate a jet from a hard lepton
or another hadronic jet. However, the sequential recombination algorithms
generally do not accommodate varying choices of radii on a jet-by-jet basis in a
single event since they have a single constant parameter that determines the
radius of a jet. Separate classifiers for a `narrow' jet and a `fat' jet in a
single event in the current $\kt$-type algorithms are thus difficult to set. An
important improvement over the current fixed radius algorithms would be to make
them adapt the jet radii dynamically jet-by-jet in each event. We make an
attempt in this direction in this work.

Our central idea of choosing the radius dynamically of a jet, especially for a
boosted fat jet, is based on the kinematics of the decay products of the
initiating heavy particle. From the theoretical side, the formation of boosted
fat jet occurs due to the high collimation of the on-shell decay products -- and
their showering  and subsequent hadronization -- of the energetic and therefore
boosted heavy particles. This is very different from the formation of light
quark- or gluon-initiated jets, whose collimation is primarily due to parton
showering and subsequent hadronization. On the other hand, at the operational
level, as per the standard $\kt$-type algorithms, the fat jets are formed in the
same way as the regular `narrow' jets, which are initiated by light quarks or
gluons. However, the kinematics of on-shell decay products and their radiation
pattern of a heavy particle is different from the showering of energetic light
quarks or gluons. Therefore, the internal structure of a fat jet is very
different from a narrow one. These internal structure has been used to tag
different heavy and light jets in the LHC context. For example, jet substructure
(JSS) observable generalized angularities
$\lambda_\beta^\kappa$\,\cite{Larkoski:2014pca,Gras:2017jty} is used to
distinguish between quark- and gluon-initiated
jets\,\cite{Gallicchio:2011xq,Gallicchio:2012ez,Andersen:2016qtm,Bhattacherjee:2016bpy,Komiske:2018vkc,Bright-Thonney:2018mxq,Larkoski:2019nwj,Nayak:2019quy,Reichelt:2021svh,Stewart:2022ari,Bright-Thonney:2022xkx}.
The same variable was used in the classification among the narrow jet, fat $W$
jet, or boosted top
jet\,\cite{Ellis:2012sn,Ellis:2014eya,Proceedings:2018jsb,Marzani:2019hun}.
Another important set of JSS observables, namely the energy correlation
functions (ECFs)\,\cite{Larkoski:2013eya,Larkoski:2015kga}, was shown to be
useful in classifying different types of
jets\,\cite{Larkoski:2014gra,Larkoski:2014zma,Adams:2015hiv,Moult:2016cvt,Proceedings:2018jsb,Chivukula:2017nvl}.
The observable N-subjettiness ($\tau_N$)\,\cite{Thaler:2010tr,Stewart:2015waa}
has been used to find the multi-pronged nature of light or heavy
jets\,\cite{Dey:2021sug,Thaler:2011gf,Han:2011ab,Plehn:2011tg,Kasieczka:2015jma,Dasgupta:2015lxh,Salam:2016yht,Han:2017hyv,Aguilar-Saavedra:2017rzt,Das:2017gke,Moore:2018lsr,Napoletano:2018ohv,Agashe:2018leo,Bradshaw:2019ipy,Aguilar-Saavedra:2020sxp,Mehtar-Tani:2020oux,Bhardwaj:2020llc,De:2020iwq,Bhattacharya:2020aid,Choudhury:2021nib,Dasgupta:2021kgi,Lu:2022cxg,Bhardwaj:2022wfz,Dey:2022tbp,Bhattacherjee:2022gjq,Das:2023owa}.
These variables have also been used extensively by the experimental
collaborations at the 13~TeV LHC\,\cite{CMS:2018ypj,ATLAS:2019kwg,CMS:2021iwu}.
These examples try to exploit the energy distribution pattern inside a jet to
distinguish a heavy object from a QCD jet. The common theme of these jet
substructure variables is the utilization of the `multi-pronged' nature of the
fat jets. Due to this multi-pronged nature, one expects the variance of
inter-constituent distance $\dR$ of a fat jet to be significantly different
compared to the narrow QCD jets. This variance of a jet can be used to grow the
radius of a jet starting from an initial radius. Earlier attempts to make the
jet radius variable, albeit with somewhat different motivations and formalisms,
can be found in references\,\cite{Krohn:2009zg,Mackey:2015hwa}. In
Ref.\,\cite{Krohn:2009zg}, the effective radius of a pseudojet during their
evolution was taken to be inversely proportional to the $\pt$ with a maximum
cut-off on the radius. Essentially, this algorithm starts from a big effective
radius and the size shrinks as a process of evolution. On the other hand, in
Ref.\,\cite{Mackey:2015hwa}, an expectation-maximization approach was taken for
clustering the hadrons into a pre-determined number of clusters (jets). Our
approach, in this work, is to modify the standard fixed radius $\kt$-type
algorithms to make the radius grow depending on the local kinematics and
distribution (in the $\eta$-$\phi$ plane) of the hadrons.

The rest of the article is organized as follows. In section 2, we briefly
outline the $\kt$-type sequential recombination algorithms followed by our
improvement to the same. We test the efficacy of our algorithms on two SM
processes and discuss them in section 3. Section 4 deals with one
application in the BSM scenario. We summarize and conclude in section~5.

%%%%%%%%%%%%%%%%%%%%%%%%%%%%%%%%%%%%%%%%%%%%%%%%%%%%%%%%%%%%%%%%%%%%%%%%%%%%%%%%
\section{Methodology}\label{sec:method}
\subsection{Standard Sequential Recombination Algorithms}\label{sec:seqalgo}
At the operational level, a jet is constituted by a bunch of four-momenta
obtained using some clustering algorithm. Among various possible ways of
grouping up the four-momenta of an event, we need to choose those relevant to
physics at the collider. It is important that the clustering algorithm should
ensure infrared and collinear (IRC) safety, which, in our context, can be
defined in terms of the following conditions\,\cite{Salam:2010nqg}:
\begin{description}
    \item [Infrared (IR) safety:] The output of the algorithm should not be
    affected by the introduction of a four-momentum with $p \to 0$.
    \item [Collinear (C) safety:] The output of the algorithm should not be
    affected by a collinear splitting of any four-momentum.
\end{description}
The algorithm that best takes care of the issue of IRC safety is known as
$\kt$-type sequential recombination jet clustering
algorithms\,\cite{Salam:2010nqg}. We briefly outline these algorithms
below\footnote{Here, we only discuss the inclusive algorithms in the LHC
context. For other jet clustering algorithms, please see
Ref.\,\cite{Salam:2010nqg}.}.

If an event consists of $N$ final state particles, whose four-momenta are taken
in a list as an input of the $\kt$-type algorithms. The distance $\dij$ between
the $\ith$ and $\jth$ four-momenta and the distance $\diB$ between the $\ith$
and the beam are then defined as
\begin{eqnarray}
\dij &=& \min\left(\pti^{2p},\ \ptj^{2p}\right) \dRij^2, \label{eqn:dij}\\
\diB &=& \pti^{2p} \Rnot^2, \label{eqn:diB}
\end{eqnarray}
where $\Rnot$ is the radius parameter of the algorithm, $\dRij$ is the Euclidean
distance between the $\ith$ and $\jth$ four-momenta in the $\eta$-$\phi$ plane,
and $\pti$ is $\pt$ of $\ith$ four-momenta. The exponent $p$ sets the weight
factor to the Euclidean distance in the $\eta$-$\phi$ plane. The three choices
of $p=1$, 0 and $-1$ correspond to the
$\kt$ (KT)\,\cite{Catani:1991hj,Catani:1993hr,Ellis:1993tq},
Cambridge-Aachen (CA)\,\cite{Wobisch:1998wt,Dokshitzer:1997in}, and
anti-$\kt$ (AK)\,\cite{Cacciari:2008gp} algorithms, respectively. The algorithm
for combining nearby four-momenta with respect to the above distance measures to
form jets has the following steps.
\begin{description}
    \item [Step 1.] The distances $\dij$ for all the possible pairs and beam
    distances $\diB$ for all the four-momenta are calculated first.

    \item [Step 2.] The minimum among all the $\dij$ and $\diB$'s is determined.

    \item [Step 3a.] If the minimum occurs at one of the $i,j$ pairs, the
    corresponding $\ith$ and $\jth$ four-momenta are merged to form a new
    four-momentum. The older ones, $\ith$ and $\jth$ four-momenta are removed
    from the list and the newly merged one is added to the list and goes back to
    Step~1.

    \item [Step 3b.] On the other hand, if the minimum distance is one of the
    $\diB$, the $\ith$ four-momenta is declared as a final jet, and it is
    removed from the list and goes back to Step~1.

    \item [Step 4.] The process is stopped once the list gets empty.
\end{description} 

This class of algorithms is seedless because the clustering of four-momenta to
form a jet does not start from a particular seed. Rather, the algorithms try to
merge the closest pair first. A group of hadrons is then declared as a jet when
an appropriate size is reached. The essential difference among the three
different algorithms, {\it viz.} AK, CA, and KT is that they give different
weights to the Euclidean distance in the $\eta$-$\phi$ plane. This typically
sets some sort of seed to the clustering algorithms in the sense that it gives a
preference to a hadron around which four-momenta merge to give rise to a final
jet. In the case of the KT algorithm, it is the softer (in terms of $\pt$)
constituent which merges first and then the harder ones get attached to it. As a
result, the shape of the final jet may not be circular in the $\eta$-$\phi$
plane. On the other hand, in the AK algorithm, the hardest particle in a
neighbourhood becomes some sort of seed for the jet and the softer ones merge at
a later stage. Hence the final jet looks circular in the $\eta$-$\phi$ plane. In
the CA algorithm, the merging is purely angular. Among the three algorithms, the
AK algorithm is the most popular one owing to its circular shape. Importantly,
in the $\kt$-type algorithms, there is a fixed radius parameter $\Rnot$, whose
value dictates the typical size of all the jets in a particular event. We note
that these algorithms are unable to capture the essential features of the events
where narrow and fat jets may simultaneously arise. In our proposed algorithm,
we have modified these algorithms to bring out the features of varying sizes of
the jets.

\subsection{Our Proposal: Dynamic Radius Jet Clustering Algorithm}
The usual $k_t$-type algorithms take a fixed radius as an input parameter, and
hence the algorithms return all the jets to be of the same size (or narrower) in
a single event. This lack of dynamicity in choosing a radius can be overcome by
setting the radius parameter dynamically during the construction of each jet.

In any $\kt$-type algorithm, the starting point is a list of $N$ four-momenta of
particles. We will refer to these as fundamental particles or, sometimes,
fundamental four-momenta. The algorithm follows Steps 1 to 3b, as defined in
section~\ref{sec:seqalgo}, iteratively until the list gets empty. At every
iteration, the number of contents of the list gets reduced by one. The reduction
happens in two ways: (1) via the merger of two four-momenta, (2) via the
declaration of four-momentum as a final jet. Thus at an intermediate iteration,
the list contains two different types of objects. These two types of objects are
(1) fundamental four-momenta, and (2) composite four-momenta, generated through
the merger of two or more fundamental four-momenta. These composite objects
evolve through iterations to give rise to the final jets. For our convenience,
let us label these composite evolving objects as pseudojets. We borrowed the
name pseudojet from the {\tt PseudoJet} class in the {\tt FastJet3}
package\,\cite{Cacciari:2011ma}, where all the types of four-momenta are called
pseudojet. However, we will call them by different names: fundamental, pseudojet
(composite or evolving), and jet (or final jet).

Our proposal is to change the constant nature of the radius parameter $\Rnot$ in
Eq.~(\ref{eqn:diB}) to a dynamic quantity depending on the distribution, in the
$\eta$-$\phi$ plane, of the fundamental objects inside each evolving pseudojet.
Therefore, the modified distance measure for the dynamic radius algorithm takes
the form
\begin{eqnarray}
\dij &=& \min\left(\pti^{2p},\ \ptj^{2p}\right) \dRij^2 \label{eqn:dijp},\\
\diB &=& \pti^{2p}\,\Rdi^2\label{eqn:diBp},
\end{eqnarray}
where $\Rdi$ is the dynamical radius parameter, defined as
\begin{eqnarray}
\Rdi = \Rnot + \sigmaRi. \label{eqn:Rdij}
\end{eqnarray}
The constant $\Rnot$ is an input parameter similar to the standard $\kt$-type
algorithm and it is the starting point of the dynamical growth of the radius of
an evolving jet. For the $\ith$ pseudojet, $\sigmaRi$ is calculated as
\begin{eqnarray}
\sigmaRi^2 = \dfrac{\sumab \pta\,\ptb\ \dRab^2 }
{\sumab \pta\,\ptb } - \left(\dfrac{\sumab \pta\,\ptb\ \dRab}
{\sumab \pta\,\ptb }\right)^2, \label{eqn:var}
\end{eqnarray}
where the summation indices $a$ and $b$ run over the fundamental constituents of
the pseudojet. The modifier $\sigmaRi$ of the radius parameter in
Eq.~(\ref{eqn:var}) is basically `$\pt$-weighted' standard deviation of the
distances between pairs of fundamental constituents of an evolving pseudojet. In
our proposal, this standard deviation $\sigmaRi$ is used to capture the size
feature of an evolving jet dynamically. For a single fundamental four-momentum,
$\sigmaRi$ is taken to be zero.

The motivation for choosing the modifier of the radius parameter to be
$\pt$-weighted standard deviation is as follows. As more than one fundamental
objects merge to become a new pseudojet, it no longer represents a single point
in the $\eta$-$\phi$ plane; it is a composite object whose constituents are
distributed in that plane. The standard deviation $\sigmaRi$ for a pseudojet
$i$, defined in Eq.~(\ref{eqn:var}), provides a measure of its fuzziness. We
want to incorporate this fuzziness in the radius parameter. In the measure of
its fuzziness, we also want the harder components to be more dominant than the
softer ones. Essentially, if the pseudojet is dominated by a single $\pt$-hard
fundamental constituent or many extremely collimated but similar $\pt$ objects,
we do not want its radius to get increased further. This is because, in these
scenarios, the final jet is expected to be a narrow jet. On the other hand, if
the pseudojet has more than one $\pt$-hard fundamental constituents slightly
separated, we expect it to be a fat jet and therefore need an increment to its
radius. Both of these two aspects are taken care of by the $\pt$-weighted
standard deviation in Eq.~(\ref{eqn:var}).

Thus, in our proposal, we first take a starting radius $\Rnot$ to be our input
parameter. The algorithm then calculates $\Rdi$ for each pseudojet, which at an
intermediate state accumulates some constituents. At every iteration, the value
of the dynamic radius parameter is calculated as the sum of the starting radius
$\Rnot$ and the radius modifier $\sigmaR$. In a nutshell, the proposed algorithm
starts from an initial radius $\Rnot$ and grows its radius dynamically using the
information from the distribution of its constituents in the $\eta-\phi$ plane.

In the proposed algorithm, the exponent $p$ to the $\pt$ in the expressions of
distance measures $\dij$ and $\diB$ in Eqs.~(\ref{eqn:dijp}--\ref{eqn:diBp}) can
take three possible values. We will call the corresponding algorithms as dynamic
radius AK (DR-AK), dynamic radius CA (DR-CA), and dynamic radius KT (DR-KT) jet
clustering algorithms.

The IRC safety of the algorithm through the definitions provided in
section~\ref{sec:seqalgo} can be approximately ensured in the radius modifier
$\sigmaR$ as well as in the final output of the algorithm. With the introduction
of an additional four-momentum, say $\pnew$, the additive contributions to the
numerators and to the denominators of the two terms in Eq.~(\ref{eqn:var}) can
be generically written as $\ptnew\suma\big(\pta\dRanew^\alpha\big)$ (for the
denominators, $\alpha = 0$, and for the two numerators $\alpha = 1$ and 2).
Clearly, all the additive contributions go to zero as $\ptnew\to 0$, thereby
ensuring the IR safety of the quantity $\sigmaRi$ for $\ith$ pseudojet. For the
consideration of IR safety of the algorithm, let us assume an extra particle of
momentum $\pnew$ is introduced in an existing event. This extra particle
actively participates in the clustering process only by one of the three
actions: (a) by getting merged to another fundamental particle, (b) by getting
recombined to a composite pseudojet, or (c) by getting declared as a singleton
jet. The action (a) does not change the value of $\sigmaR$ or the four-momentum
of the pseudojet after the merger of the four-momentum with $\pnew\to 0$. The
same is true for the merger of $\pnew$ via action (b) since the merger of two
fundamental four-momenta keeps the value of $\sigmaR$ at zero. After the merger
of this $p\to 0$ four-momentum, both the radius modifier $\sigmaR$ and the total
momentum remain unaffected. After this merger, the rest of the clustering
process does not get affected, and hence the final output of the clustering
algorithm remains unaffected. Furthermore, action (c) does not give rise to an
extra jet whenever $p\to 0$.

For the collinear safety, one can see that the radius modifier $\sigmaR$ remains
almost unaltered when a four-momentum is split collinearly. Let a four-momentum
$\pnew$ gets split into $\pr$ and $\ps$. Any general term
$\pta\ptnew\dRanew^\alpha$ then becomes $\pta(\ptr\dRar^\alpha +
\pts\dRas^\alpha)$. In the collinear splitting limit, $\ptnew = \ptr+\pts$,
$\dRar=\dRas=\dRanew$. Moreover, there will not be any additional contribution
due to the $\pr$ and $\ps$ combination except for the denominators in
Eq.~(\ref{eqn:var}) since $\dRrs=0$. This ensures an approximate collinear
safety of $\sigmaRi$ for any $\ith$ pseudojet. On the other hand, for the
collinear safety of the algorithm, if any four-momentum collinearly splits into
two four-momenta, then the distance $\dij\to 0$. Hence, these two collinearly
split four-momenta get merged together at a very early stage; a feature that is
inherently present in the $\kt$-type algorithm. Other IRC safety features (due
to $\pt$-dependent prefactors in the $\dij$ and $\diB$ definitions) of the
standard $\kt$-type algorithms will be inherited by the dynamic radius
algorithm.

We have implemented the method of dynamic radius jet clustering algorithm as a
{\tt FastJet3} plug-in\,\cite{Cacciari:2005hq,Cacciari:2011ma}. This package has
many built-in data-types and functionalities to optimize the implementation and
computation of jet clustering algorithms. In particular, we have used {\tt
NNBase} and {\tt NNH} classes to help us keep track of the distance measures. As
required by these two classes, our $\dij$ measure is also symmetric in $i$ and
$j$ indices. The {\tt ClusterSequence} class has then been used to merge two
four-momenta and keep track of the clustering sequence. The {\tt PseudoJet}
class has been used to store the four-momenta information of all the initial,
intermediate, and final jets. The {\tt user\_info} property of {\tt PseudoJet}
data-type has been used to store the information related to the radius modifier
$\sigmaRi$ of the $\ith$ pseudojet. This way of implementation has at most
$N^2$ computational complexity for an event of size $N$. The worst possible
complexity arises when all the particles in an event are merged to form a
single jet. Since this worst possibility does not generally occur, we expect
the computational expense to be less in a practical scenario. We note that the
standard $\kt$-type algorithms also have $N^2$ complexity via the basic
implementation of the FastJet algorithm\,\cite{Cacciari:2005hq,Cacciari:2011ma}.

One important point to note is that the equations for distance measures, defined
in Eqs.~(\ref{eqn:dij}--\ref{eqn:diB}), can be recast to in the radius parameter
in the expression of $\dij$ rather than in the expression of $\diB$. That is to
say that the modified set of equations can be taken to be 
\begin{eqnarray}
\tilde{\dij} &=& \min\left(\pti^{2p},\
\ptj^{2p}\right)\,\left(\frac{\dRij}{\Rnot}\right)^2 \label{eqn:dij1}\\
\tilde{\diB} &=& \pti^{2p} \label{eqn:diB1}
\end{eqnarray}
The standard sequential recombination algorithm yields identical results in both
formalisms since the radius is a constant parameter. However, if this latter
formalism is chosen to incorporate dynamicity, the form of the dynamic
radius parameter $\Rd$ will be different. The dynamic radius $\Rd$, in this type
of modification, will be dependent on both pseudojets. One option would be to
add the standard deviations $\sigmaRi$ and $\sigmaRj$ of the $\ith$ and $\jth$
pseudojets, respectively, to the constant parameter $\Rnot$. This way of
defining $\Rd$ ensures the symmetry in $i$ and $j$ indices and, therefore, the
implementation of the method via {\tt NNBase} and {\tt NNH} as a FastJet3
plug-in can easily be performed. In any case, the output of the algorithm is
modified according to Eq.~(\ref{eqn:dij1}--\ref{eqn:diB1}) will be different
from that of the one considered in Eq.~(\ref{eqn:dijp}--\ref{eqn:diBp}).

We now are ready to apply our formalism to some simple SM processes and check
how it performs compared to the standard sequential recombination algorithms. We
discuss this in the next section in connection with SM processes and consider
its application to BSM in the section after that.

%%%%%%%%%%%%%%%%%%%%%%%%%%%%%%%%%%%%%%%%%%%%%%%%%%%%%%%%%%%%%%%%%%%%%%%%%%%%%%%%
\section{Application to Standard Model Processes}\label{sec:appSM}
We take the following two SM processes to illustrate the performance of our
newly developed algorithm.
\begin{itemize}
    \item[{\bf I.}] $p p \to t j$
    \item[{\bf II.}] $p p \to V j$, ($V=W \text{ or } Z$)
\end{itemize}
For both cases, we have generated parton-level events using {\tt MadGraph5}
(MG5)\,\cite{Alwall:2014hca}. We will refer to these events as \MGpartonlevel
events and the final state partons in these events as \MGparton s. A lower cut
of 500~GeV on the $\pt$ of the jets has been imposed during the generation of
the \MGpartonlevel events. This helps us in generating events with boosted top
or vector bosons at the parton-level, which then form fat jets after subsequent
decays and hadronization. For the purpose of the following studies, only the
hadronic decays of top and $W/Z$ are considered. We have then passed the
\MGpartonlevel events to
{\tt Pythia8}\,\cite{Sjostrand:2006za,Sjostrand:2014zea} for showering and
hadronization. The {\it Monash 2013 Tune}\,\cite{Skands:2014pea}, the default
tune of Pythia8, has been used to take care of the simulations of underlying
events and multi-parton interactions in the proton-proton collisions. The output
of {\tt Pythia8} has then been transferred to {\tt FastJet3} for the formation
of jets.

%===============================================================================
\subsection{Illustration I: $pp\to tj$ process}
\begin{figure}[!h]
\begin{center}
\begin{subfigure}{1.0\textwidth}
\includegraphics[width=0.5\textwidth]{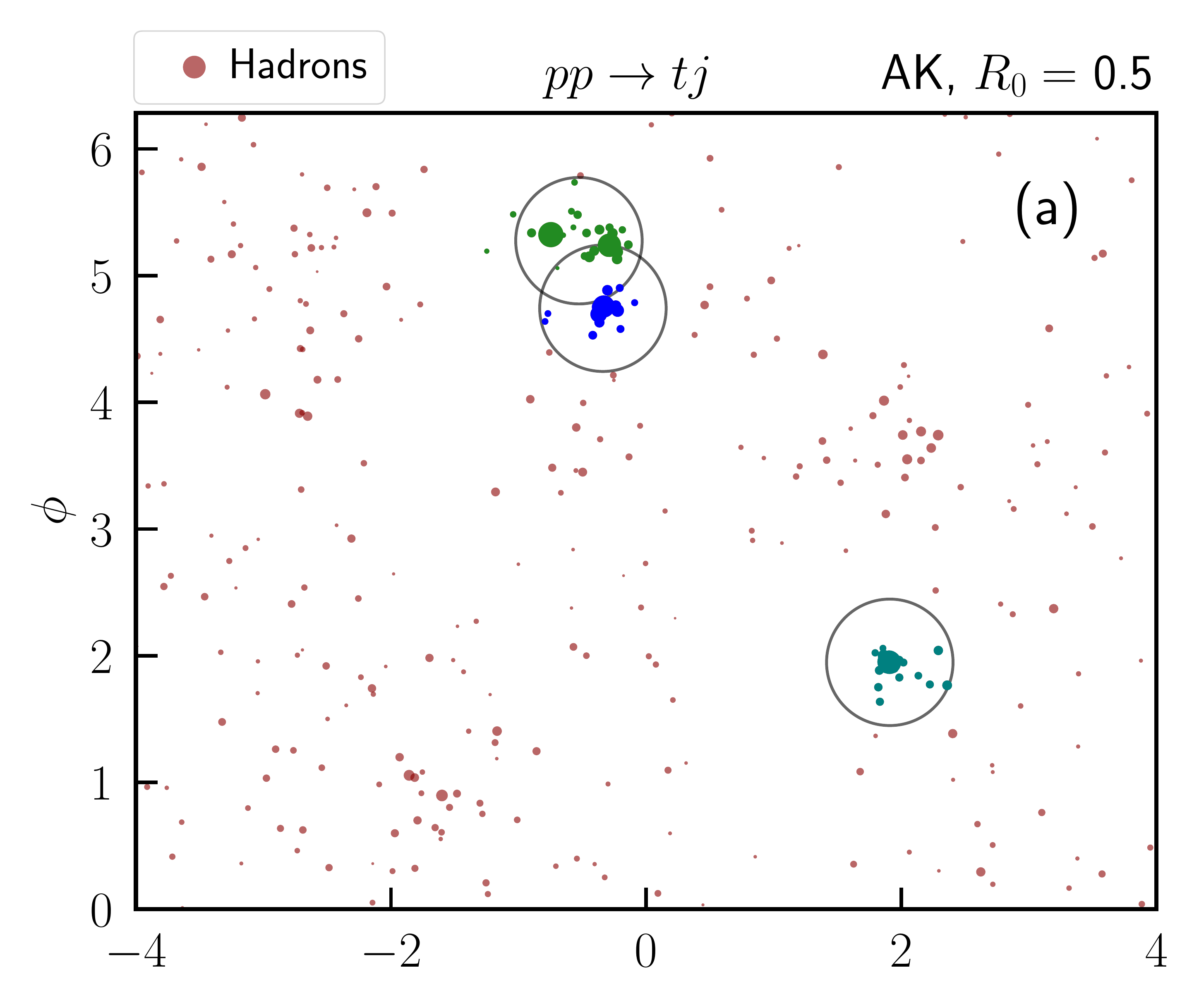}
\includegraphics[width=0.5\textwidth]{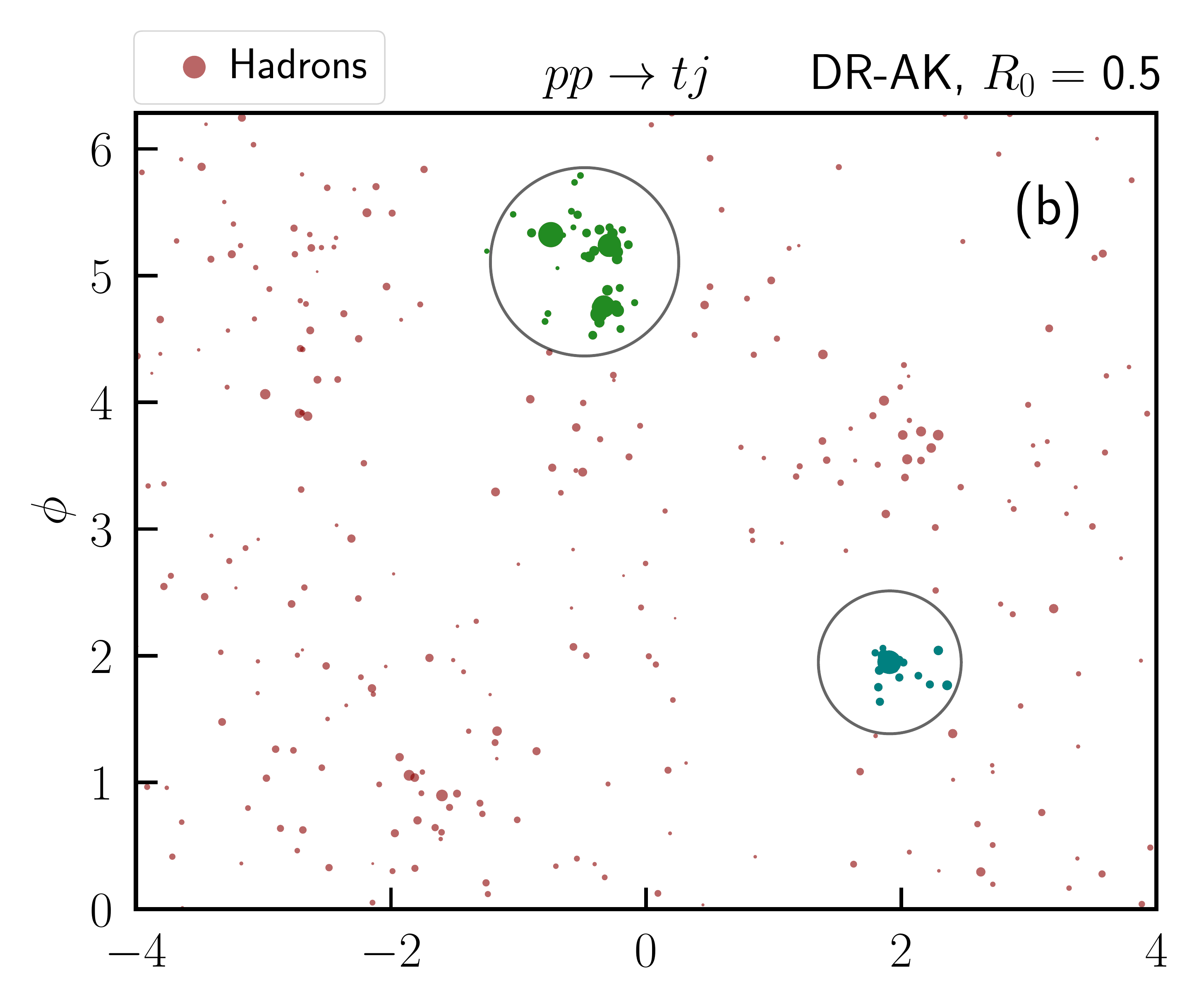}
\end{subfigure}
\begin{subfigure}{1.0\textwidth}
\includegraphics[width=0.5\textwidth]{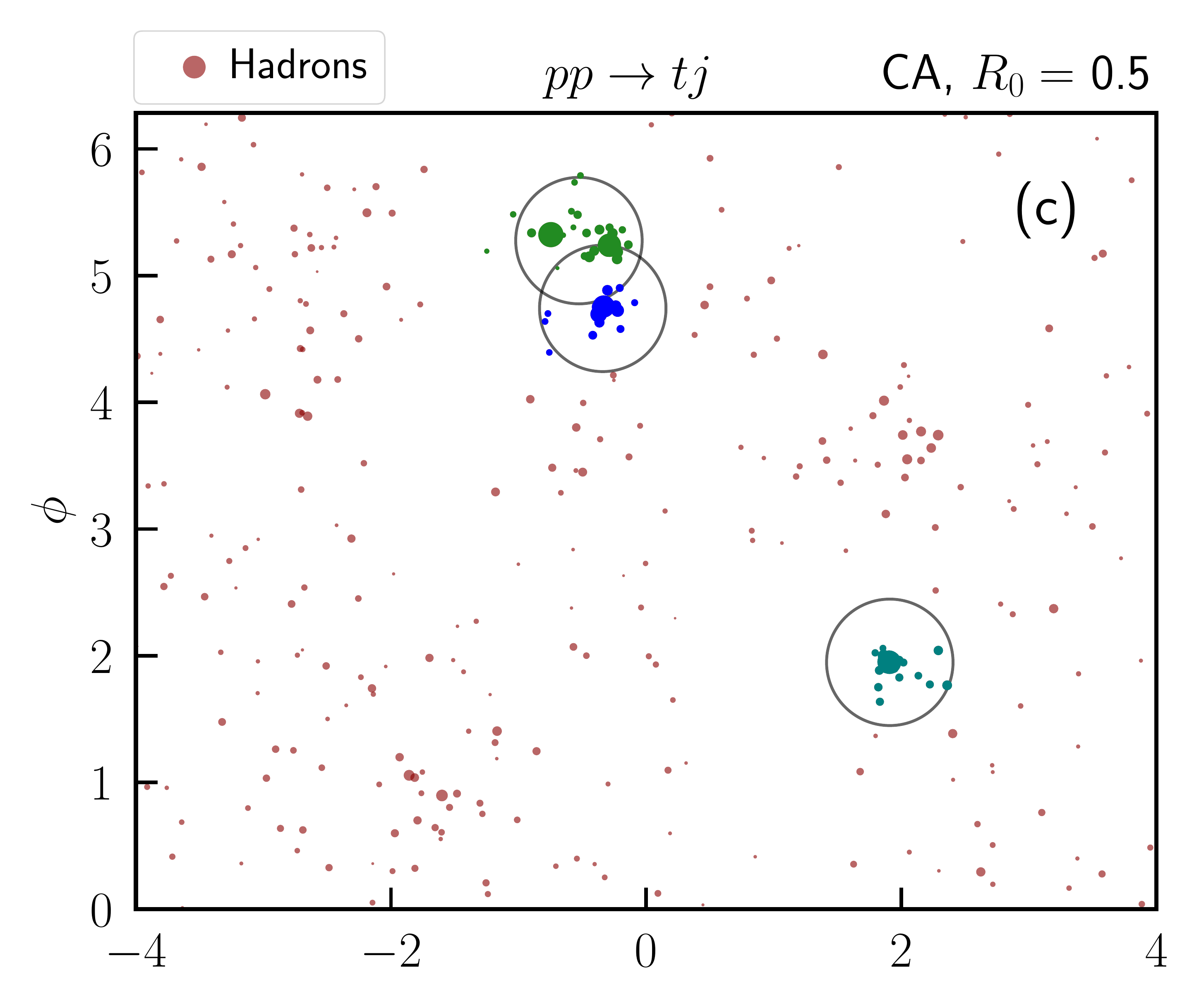}
\includegraphics[width=0.5\textwidth]{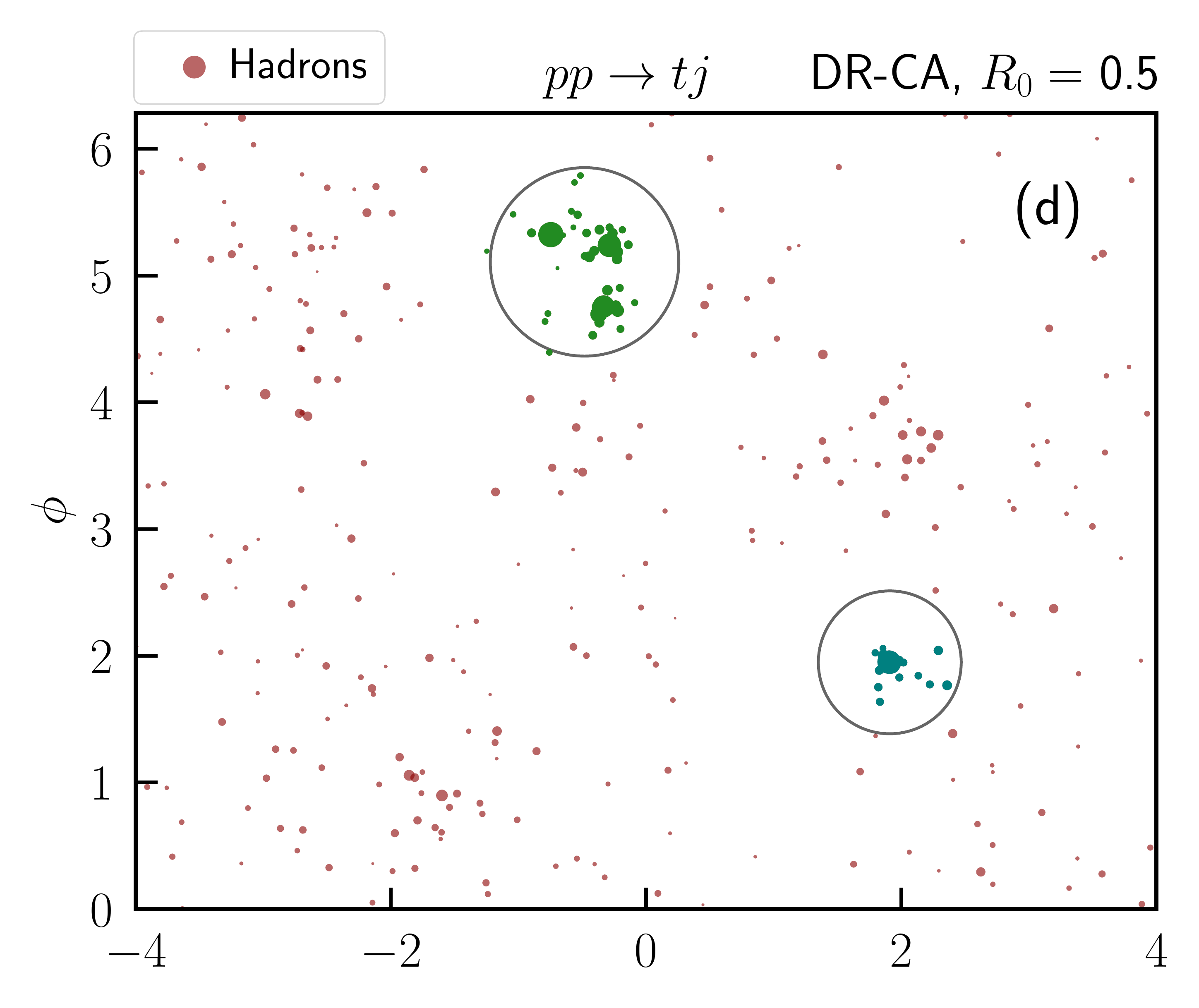}
\end{subfigure}
\begin{subfigure}{1.0\textwidth}
\includegraphics[width=0.5\textwidth]{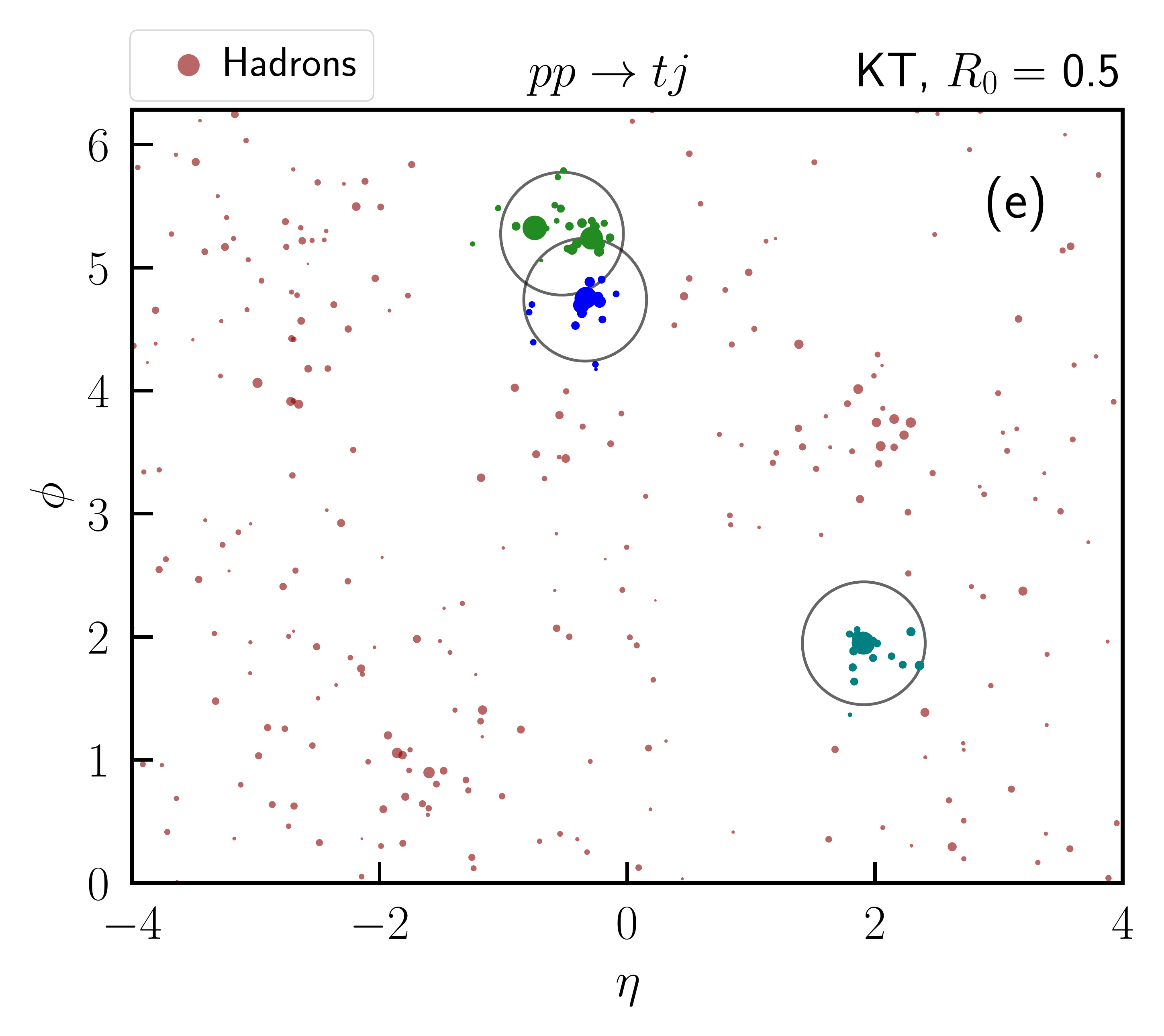}
\includegraphics[width=0.5\textwidth]{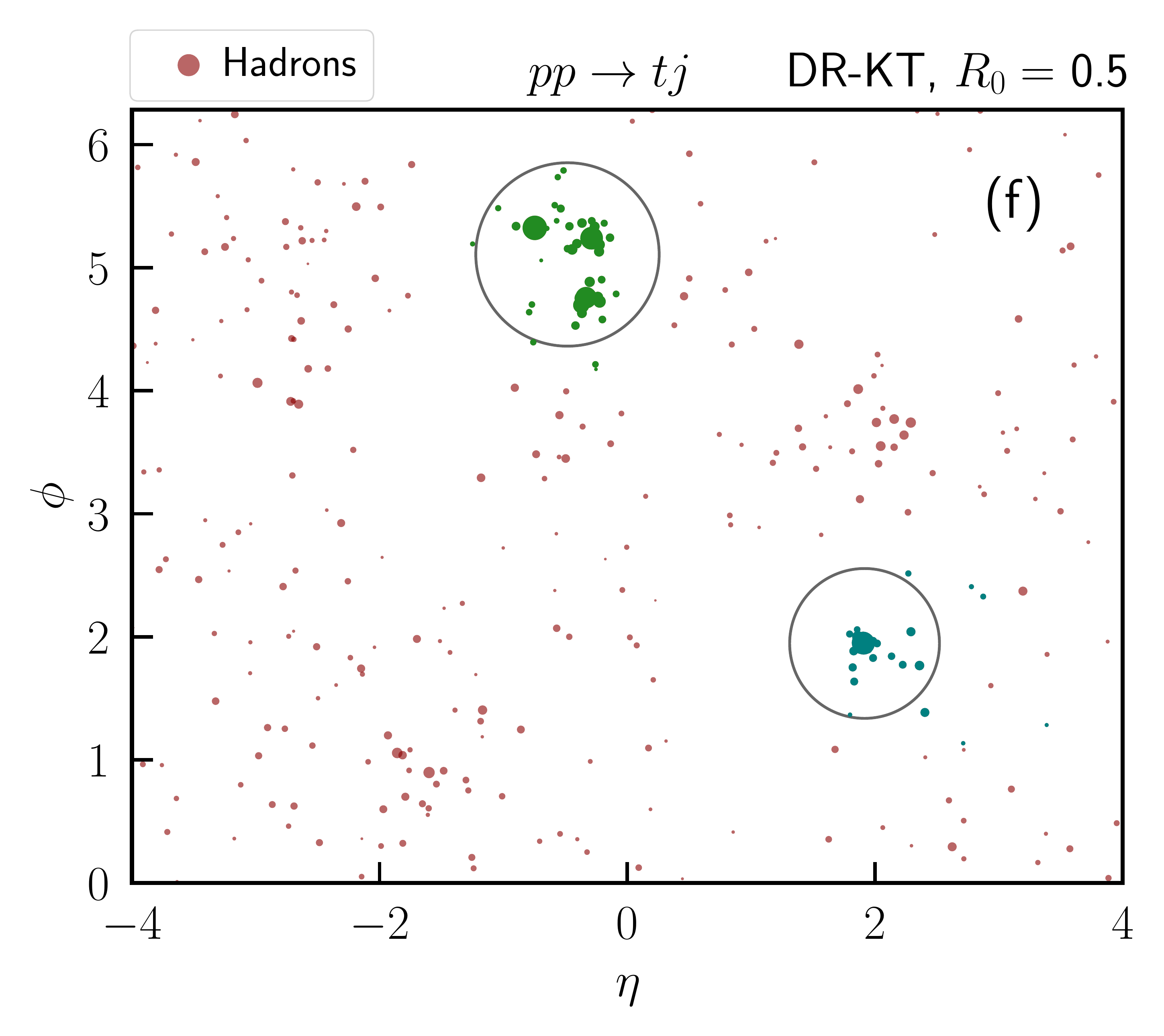}
\end{subfigure}
\caption{Positions of final state hadrons and jets in the $\eta$-$\phi$ plane in
an example event for $pp\to tj$ process. The red dots represent the final state
hadrons and their sizes are kept proportional to $\sqrt\pt$ of the corresponding
hadron. The unfilled circles represent the final radius $\Rd$ of a jet. The teal
coloured dots represent the constituents of the hard `narrow' jets. The green
and blue (wherever applicable) dots represent the constituents of the fat
\Top\ jet. The left panel, from top to bottom, is for (a) AK, (c) CA, and (e) KT
algorithms with $\Rnot=0.5$. The right panel, from top to bottom, represents
jets clustered using (b) DR-AK, (d) DR-CA, and (f) DR-KT algorithms,
respectively, with $\Rnot=0.5$.}
\label{fig:tj-example}
\end{center}
\end{figure}

The top quark, when highly boosted, results in a fat jet while the $j$ yields a
narrow jet after the effects of showering and hadronization. In order to compare
various jet properties between the dynamic radius and fixed radius algorithm, we
run these two types of algorithms on the same set of hadrons from each event. We
first demonstrate how the dynamic radius algorithms help in capturing the fat
and narrow objects in a single event. This has been demonstrated by depicting
the hadrons and jets of an example event in Fig.~\ref{fig:tj-example}, where we
plotted, in the $\eta$-$\phi$ plane, the position of the hadrons in the event
along with the high-$\pt$ jets constructed out of these hadrons. The sizes of
the dots are kept proportional to $\sqrt \pt$ of the hadrons. The jets are
represented by the unfilled black circles and the solid dots inside the black
circles comprise of the constituent hadrons of the jets. The three panels on the
left show the jets for (a) AK, (c) CA, and (e) KT jet algorithms with
$\Rnot = 0.5$. In all the left panels, the algorithms return three high-$\pt$
jets; one near (2,2) position and the other two are near (0,5) position in the
$\eta$-$\phi$ plane. With the \MGpartonlevel information, we identified that the
jet in (2,2) position is initiated by $j$ while the two jets near the (0,5)
position is initiated by the decay products of the hard top quark. Because of
fixed radii of the standard $\kt$-type jet clustering algorithms, they could not
capture all the hadrons initiated by the decay products of the top quark inside
a single jet; rather they have been split into two different jets. A quick fix
to this problem would be to increase the size of the radius parameter. This
prescription, however, ends up increasing the jet size unnecessarily, for
example at the (2,2) position where such increment is not required. This
unnecessary increase in the radius of a jet increases jet mass, especially in
the high pile-up scenario. One interesting option in such cases would be to
choose the radius according to the need of a jet. This is precisely where the
dynamic radius jet clustering algorithm is useful in this type of scenario. This
can be seen in the three panels on the right in Fig.~\ref{fig:tj-example}. There
the hadrons and the high-$\pt$ jets are drawn for dynamic radius jet clustering
algorithms with $\Rnot = 0.5$. The interesting point to note in all three
right panels is that there are two jets instead of three. The radius of the jet
near the (0,5) position has been appropriately grown to capture the full decay
products of the top quark and their radiations while the radius of the jet near
the (2,2) position did not grow much. This desirable characteristic of a jet
algorithm would be beneficial for the studies of collider events, where narrow
as well as fat jets are expected to occur simultaneously. In all the panels, the
radius of each black circle is kept to be equal to the final radius $\Rd$, as
defined in Eq.~(\ref{eqn:Rdij}), of each individual jet. For the fixed radius
jet algorithms, the final radius is essentially the fixed radius parameter
$\Rnot$.

%===============================================================================
\afterpage{\clearpage}
%===============================================================================

Fig.~\ref{fig:tj-example} gives an approximate idea of how the dynamic radius
helps us in finding a fat jet starting from a small radius. Next, we show how
often this dynamic radius jet algorithm helps us in finding the fat jet. In
order to demonstrate that, we have employed the following procedure. We first
form the jets from the hadrons and choose only the high-$\pt$ ($> 5$~GeV) jets.
We then tag the energetic jets, event by event, as reconstructed `top' or
reconstructed `jet' with the help of \MGpartonlevel information. The events are
classified into two categories, as described below.
\begin{enumerate}
    \item [\bf A1.] Category A1 consists of events satisfying the following
    conditions.
    \begin{itemize}
        \item A jet should have mass in the range (150, 200)~GeV and have $\dR
        (\text{top}^\text{truth}, \text{jet}) < 0.5$. This jet is identified as
        a reconstructed top jet. We label these reconstructed objects as
        `\topAone' in the subsequent discussions.
        \item After the tagging of the top jet, another jet should have $p_T >
        300$~GeV and should be within 0.5 distance from the original jet as
        generated by MG5. These jets are labelled as `\jetAone' in further
        discussions.
    \end{itemize}
    \item [\bf A2.] Category A2 are the events which satisfy the following
    conditions. 
    \begin{itemize}
        \item Two separate jets within 1.0 distance of the original top quark
        and having an invariant mass between 150 and 200 GeV. These two jets are
        tagged as constituent jets of the reconstructed top jet, which is a
        combination of these two constituents. These combinations are labelled
        as `\topAtwo'.
        \item Another jet having $\pt > 300$~GeV and within 0.5 radius from the
        original jet. This is labelled as `\jetAtwo'.
    \end{itemize}
\end{enumerate}

In general, any inclusive $kt$-type clustering algorithm yields as output many
soft jets along with the hard ones. The origin of these soft jets is primarily
the soft radiation due to underlying events and wide angle parton shower. These
jets are expected in both the category A1 and A2 events. Any jet having $\pt >
5$~GeV and labelled neither as \Top\ nor as \jet\ is labelled as \softjet.

The two categories have been chosen to demonstrate the usefulness of the
dynamic radius jet algorithm. Category A1 captures the whole \Top\ jet by the
jet clustering algorithm while the events in category A2 need 
post-processing after the jet clustering. Therefore, a desirable criterion of a
better-performing jet clustering algorithm would be to have more events in
category A1. In order to illustrate that, for a given category, we define
acceptance efficiency
\begin{eqnarray}\label{eqn:acc}
{\cal A} = \frac{\text{number of events accepted in a particular category}}{\text{total number of events}}.
\end{eqnarray}

%===============================================================================
%\afterpage{\clearpage}
\begin{figure}[!h]
\includegraphics[width=\textwidth]{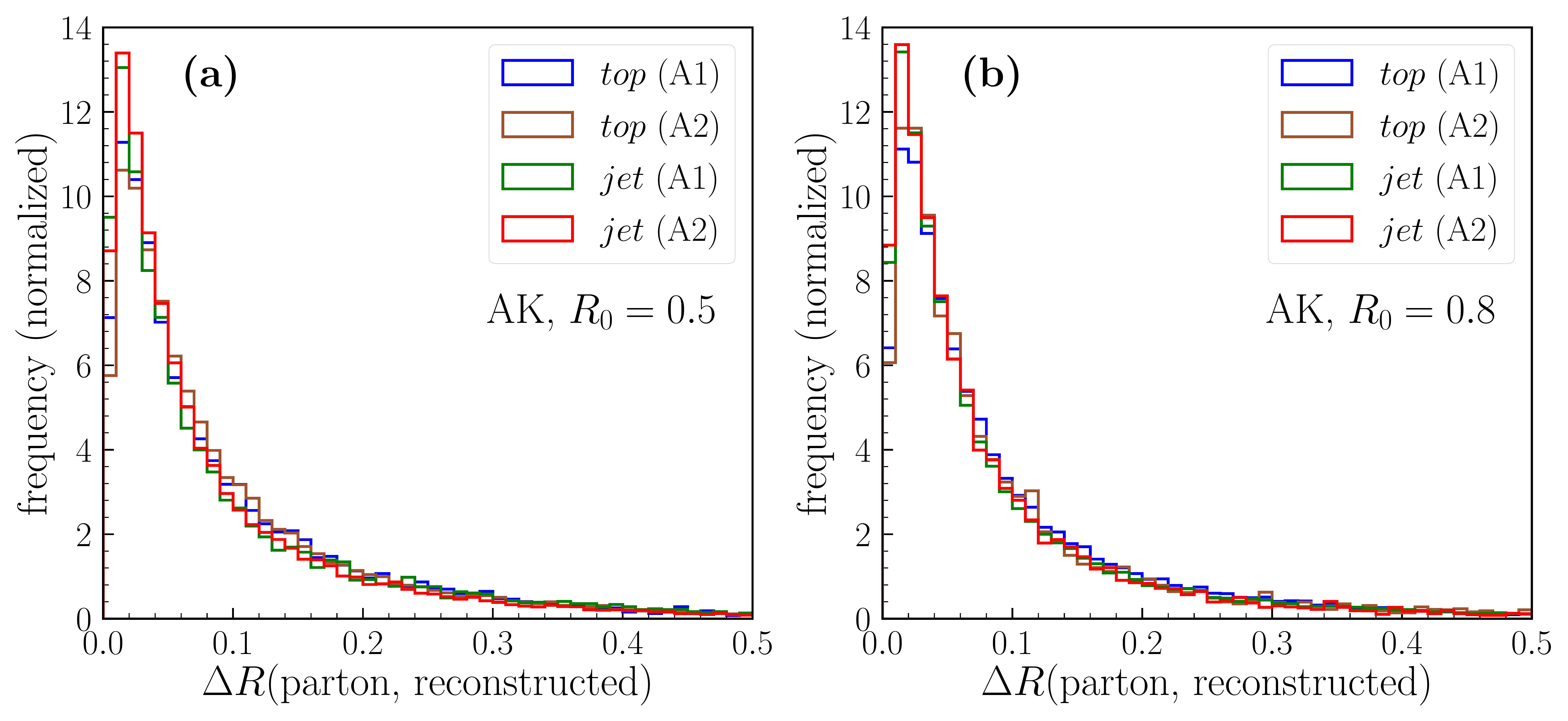}
\caption{Normalized distribution of $\dR$ between the \MGpartonlevel object and
corresponding reconstructed jet. The jets were clustered using the AK algorithm
with radius parameters (a) 0.5 and (b) 0.8.}
\label{fig:dR-tr-re}
\end{figure}
%===============================================================================

After the classification of the events into the above two categories, the
distribution of distances between the \MGpartonlevel objects and reconstructed
ones are plotted in Fig.~\ref{fig:dR-tr-re}. In both the panels of the figure,
the blue and brown histograms are for \Top\ jets, and the green and red ones are
for energetic \jets. The corresponding categories of the histograms are
mentioned alongside the legends. The distributions are shown for jets clustered
using the AK algorithm with (a) $\Rnot = 0.5$, and (b) $\Rnot = 0.8$. Since this
distance between the \MGpartonlevel and reconstructed ones are features of
parton showering and hadronization, the normalized distributions are kind of
identical for different radius choices. These $\dR$ distributions are very
similar even with different choices of standard or dynamic radius sequential
recombination algorithms and, therefore, are not shown to avoid repetition. This
distribution also justified the choice of $0.5$ radius to find reconstructed
objects from the \MGparton s.

%===============================================================================
\begin{figure}[!h]
\begin{subfigure}{\textwidth}
\includegraphics[width=\textwidth]{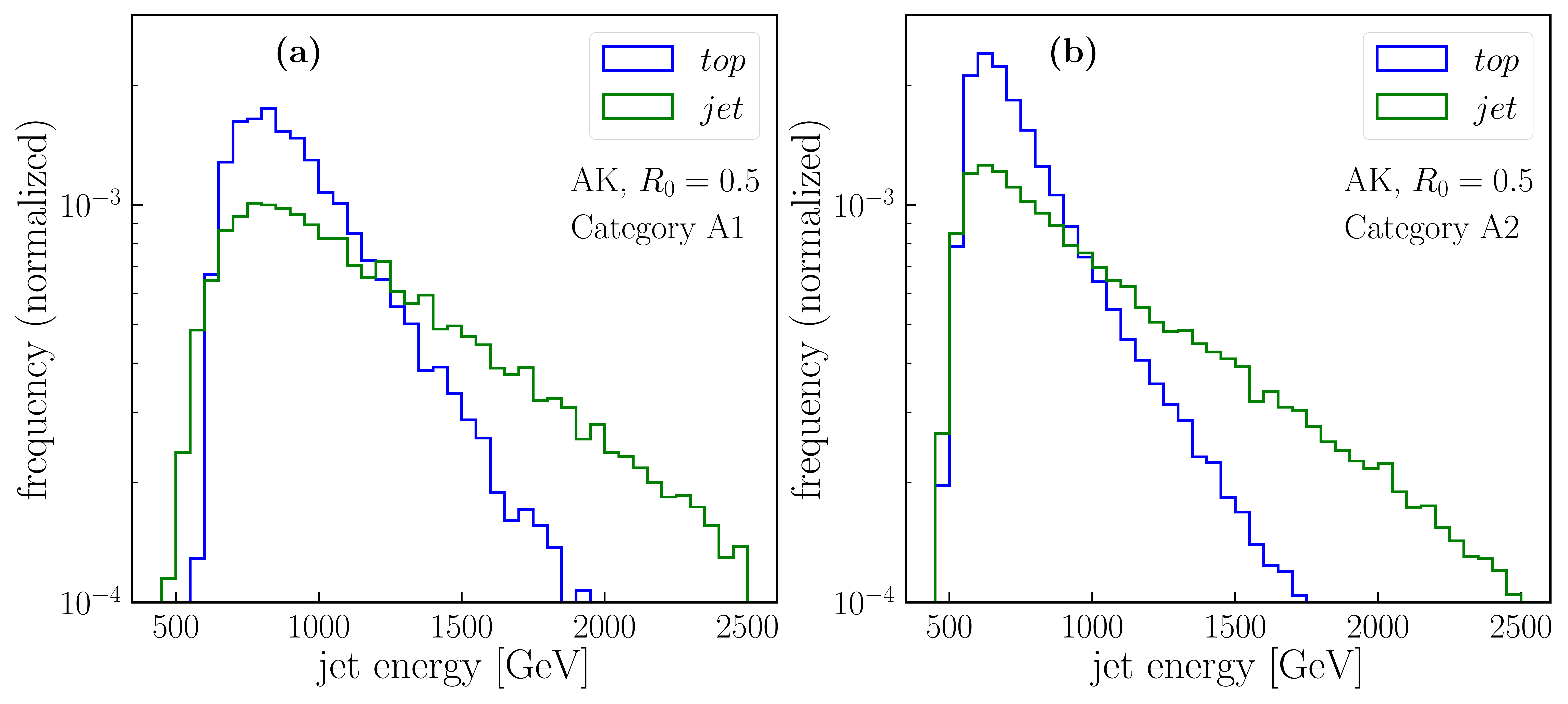}
\end{subfigure}
\caption{Normalized distribution of jet energy for categories (a) A1 and (b) A2.
The blue and green histograms are respectively for the reconstructed \Top\ and
the high-$\pt$ \jet.}
\label{fig:jetEnergy}
\end{figure}
%===============================================================================

We show in Fig.~\ref{fig:jetEnergy} the jet energy distributions for the objects
of our study. The left and right panels show the distributions for categories A1
and A2, respectively. The blue and green histograms are for the \Top\ and the
high-$\pt$ \jet\ produced in association with it.

%===============================================================================
\afterpage{\clearpage}
\begin{figure}[!h]
\begin{center}
\begin{subfigure}{\textwidth}
\includegraphics[width=0.50\textwidth]{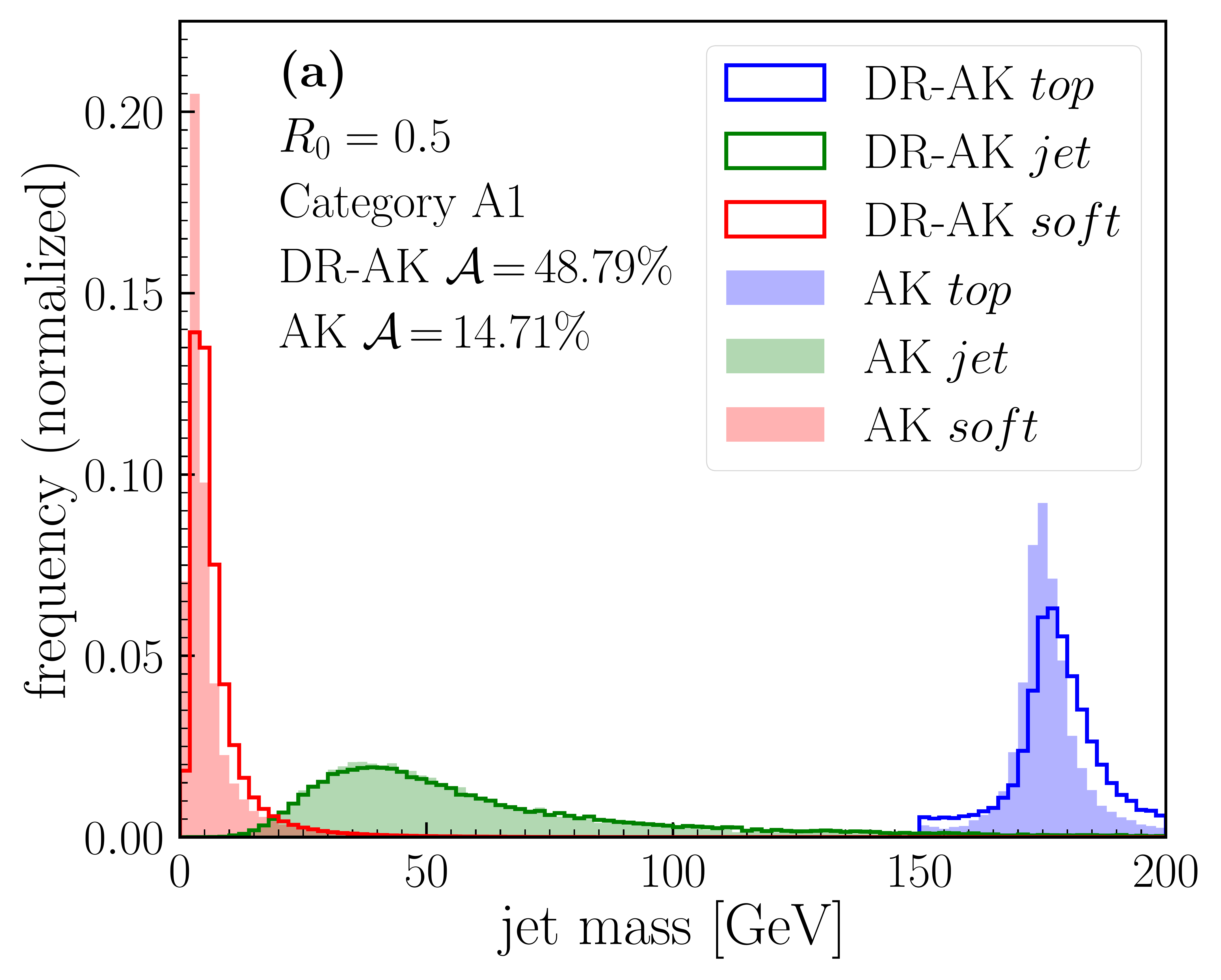}
\includegraphics[width=0.50\textwidth]{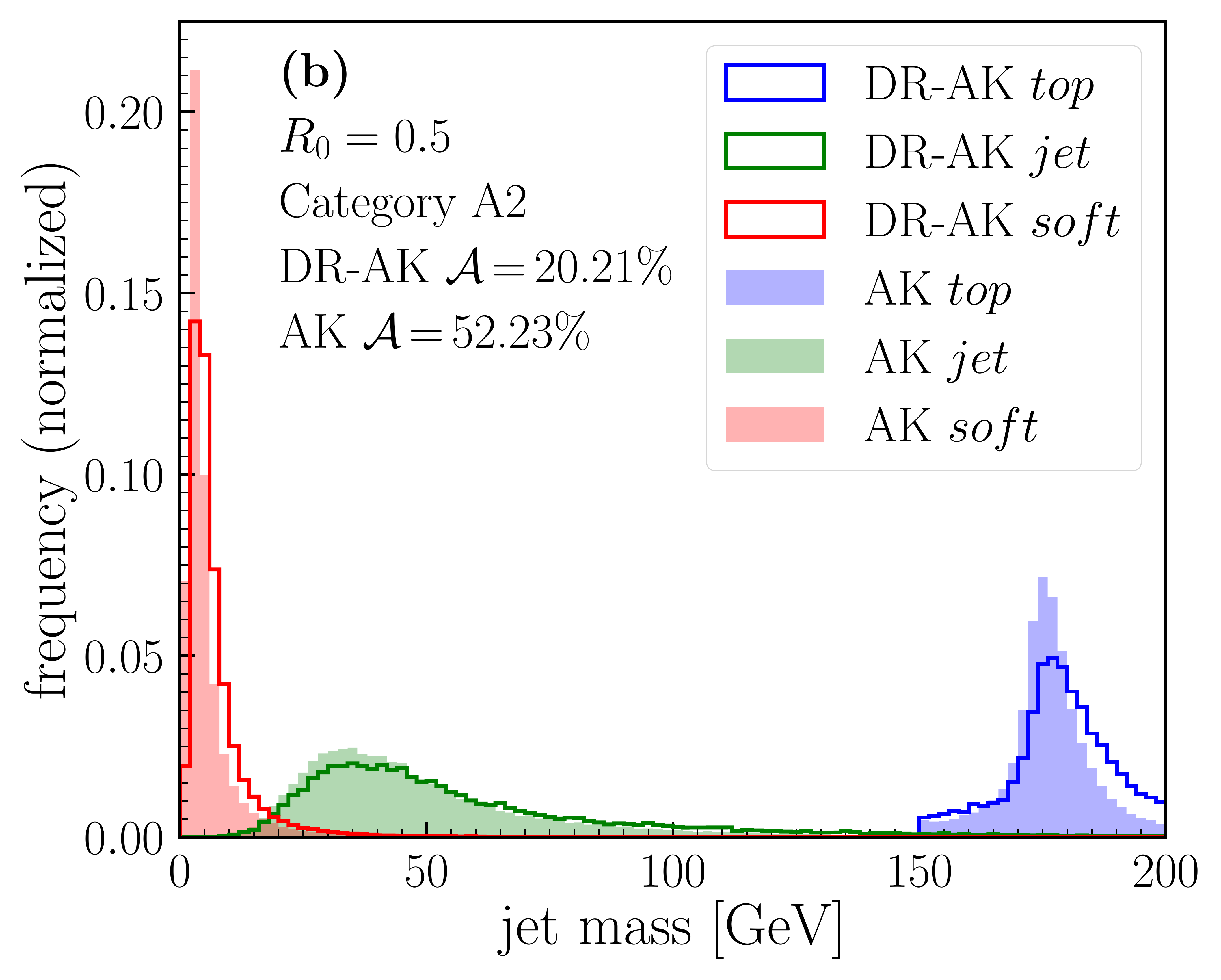}
\end{subfigure}
\begin{subfigure}{\textwidth}
\includegraphics[width=0.50\textwidth]{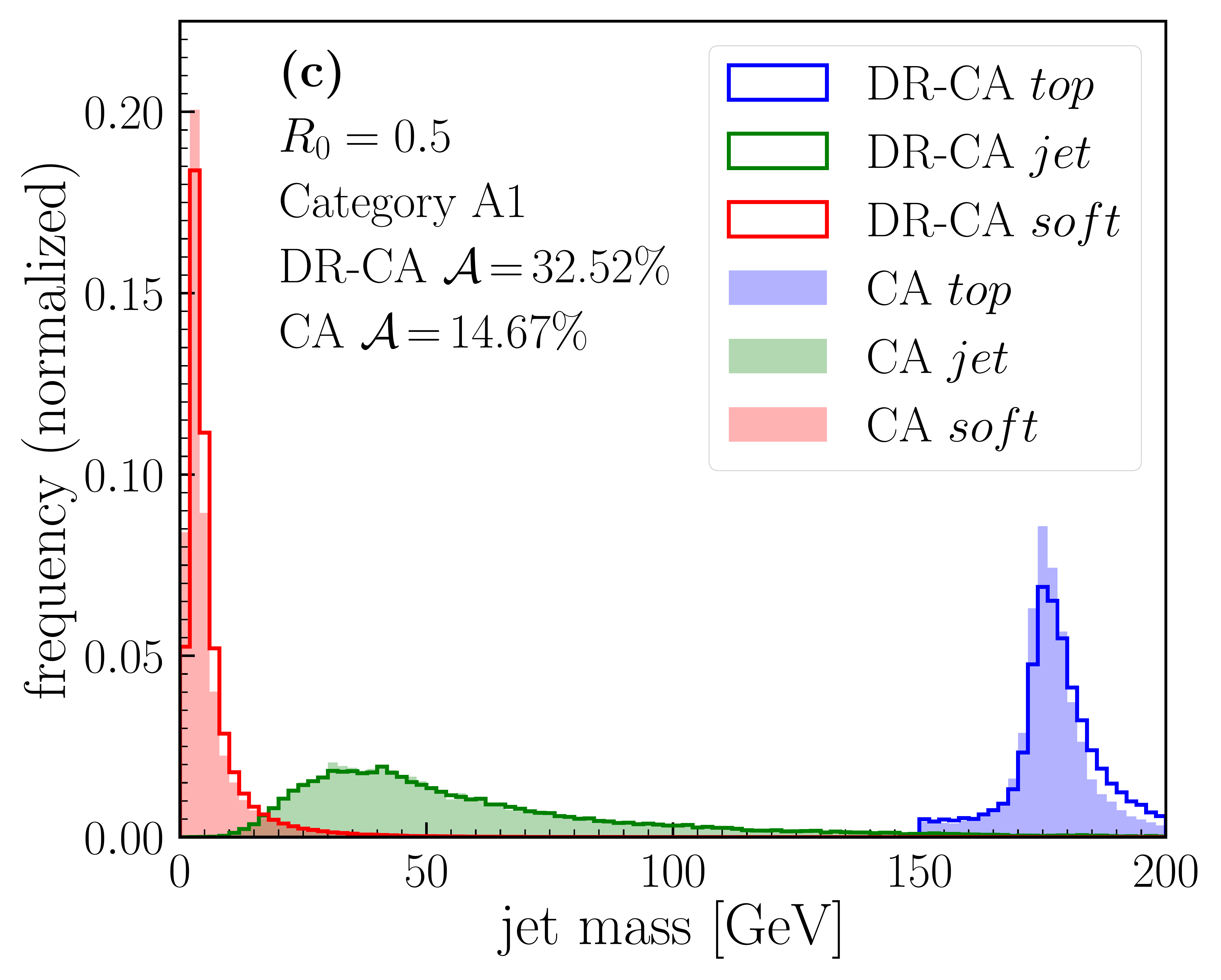}
\includegraphics[width=0.50\textwidth]{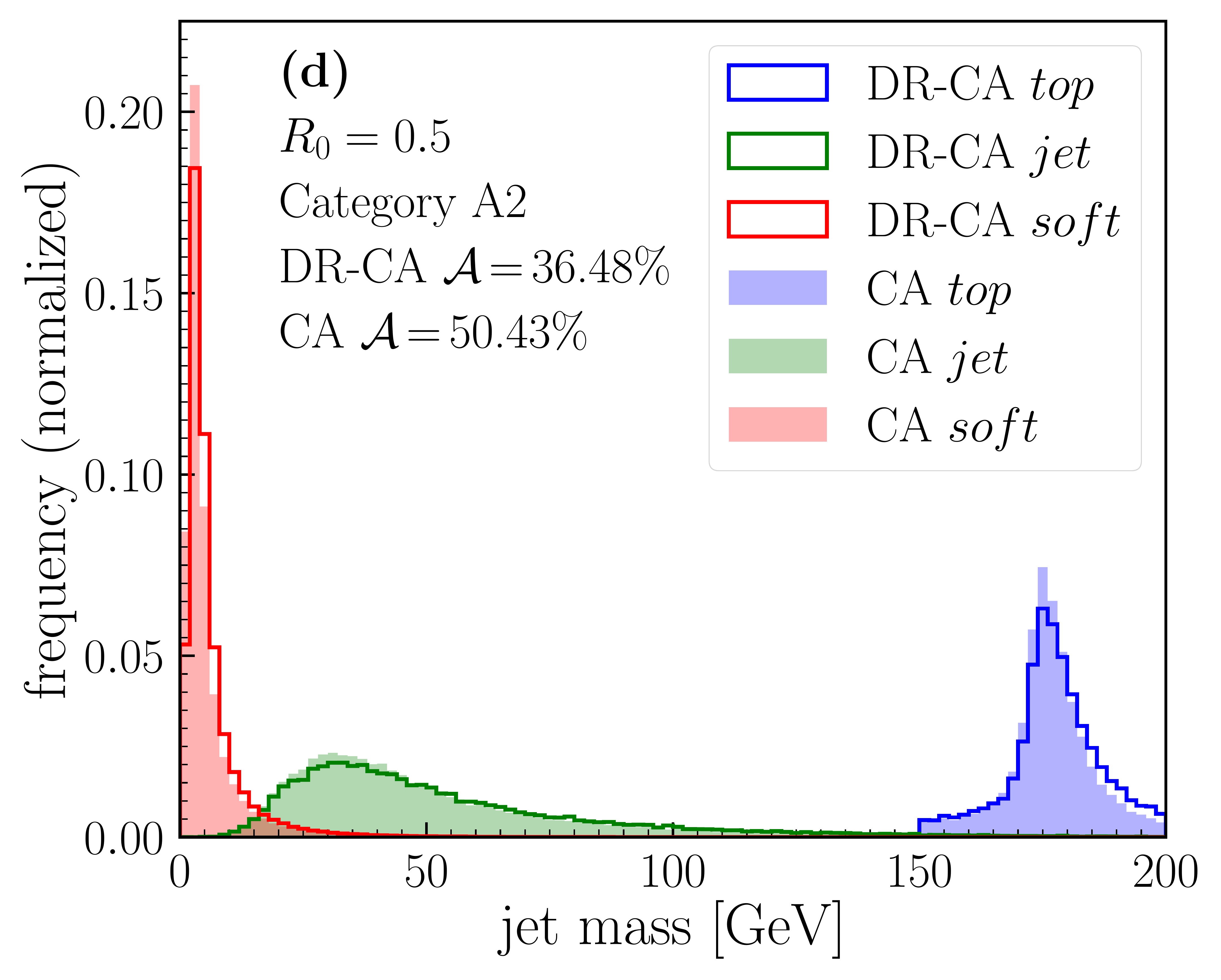}
\end{subfigure}
\begin{subfigure}{\textwidth}
\includegraphics[width=0.50\textwidth]{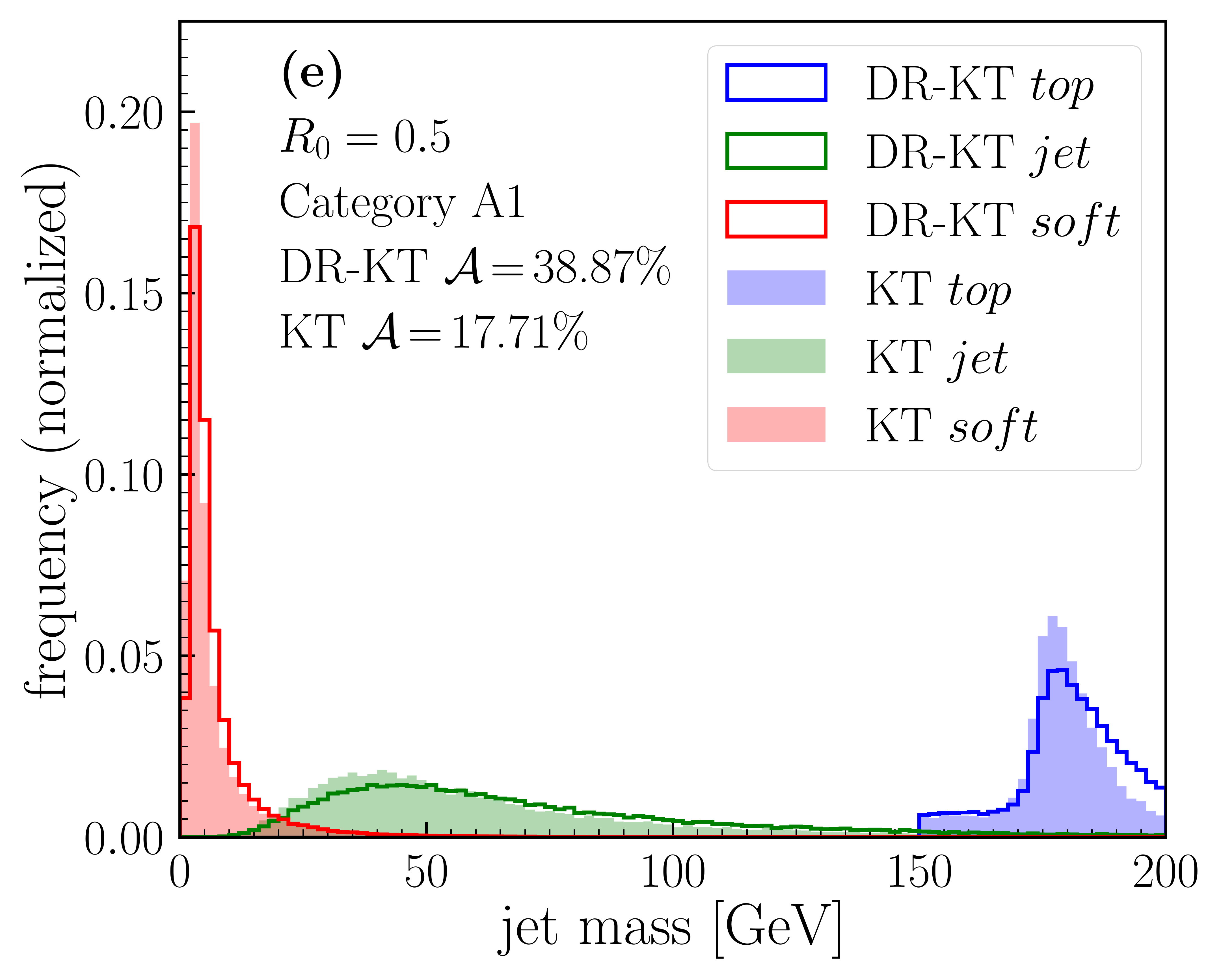}
\includegraphics[width=0.50\textwidth]{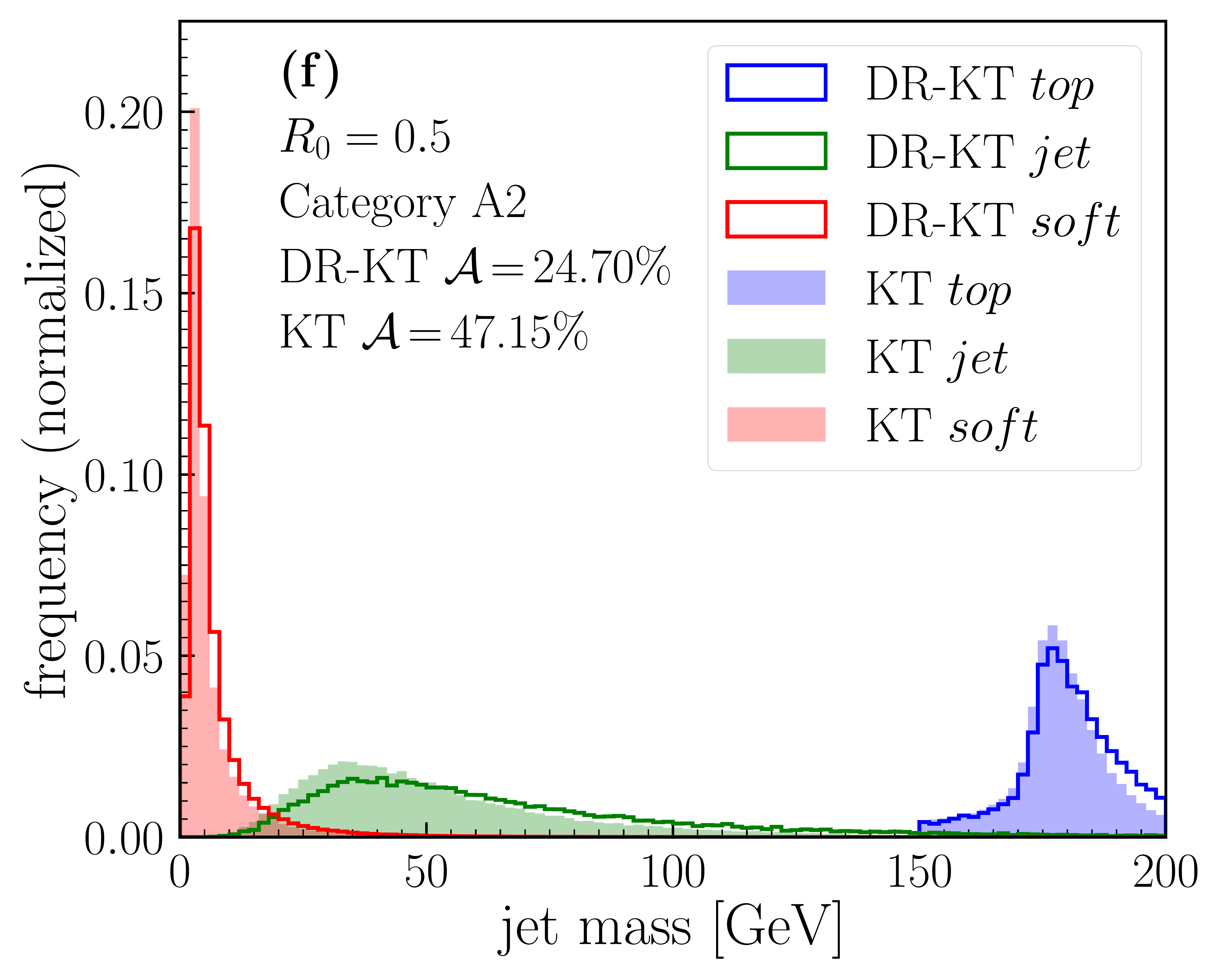}
\end{subfigure}
\end{center}%\vspace{-1cm}
\caption{Normalized distribution of jet mass for the process $pp\to t j$. The
left panel shows the distribution for category A1 events while the right panel
is the distribution for category A2 events. The blue, green, and red histograms
are for reconstructed \Top, hard \jet, and \softjets\ (defined in the text),
respectively. The histograms, from top to bottom, are for AK, CA, and KT
algorithms. The filled histograms correspond to fixed radius algorithms and the
unfilled ones correspond to their dynamic radius (DR) counterparts.}
\label{fig:pptj-hist}
\end{figure}
%===============================================================================

One of the primary obligations of choosing the appropriate size for jets
according to requirements is to avoid the rise of jet mass even with soft but
widely separated constituents inside a jet. We, therefore, choose to show the
distribution of masses of reconstructed \Top\ jets, reconstructed energetic
\jet s in Fig.~\ref{fig:pptj-hist}. The jet energy ranges corresponding to the
mass distributions shown can be approximately 500-2000~GeV, as seen in
Fig.~\ref{fig:jetEnergy}. The left panel of the figure represents the
distribution for category A1 events while the right panel represents the
distribution for category A2 events. The blue, green, and red histograms are for
reconstructed \Top, hard \jet, and \softjets, respectively. The histograms, from
top to bottom, are for anti-$\kt$, C/A, and $\kt$ algorithms. The filled
histograms are for standard jet clustering algorithms and the unfilled ones are
their dynamic radius counterparts. In the legends, the prefix `DR' to AK, CA, or
KT stands for dynamic radius. In all the panels, the starting radius parameter
has been taken to be $\Rnot = 0.5$. For standard $\kt$-type algorithms, the
starting radius is the fixed constant radius parameter, {\it i.e.}, $\Rd=\Rnot$.
The values for \Acc for different algorithms and different categories are quoted
inside each panel of Fig.~\ref{fig:pptj-hist}. In all the panels, it is seen
that the acceptance efficiencies for A1 category events in the cases with
dynamic radius algorithms are higher than their fixed radius counterparts.
 
%===============================================================================
%\afterpage{\clearpage}
\begin{figure}[!h]
\includegraphics[width=1.0\textwidth]{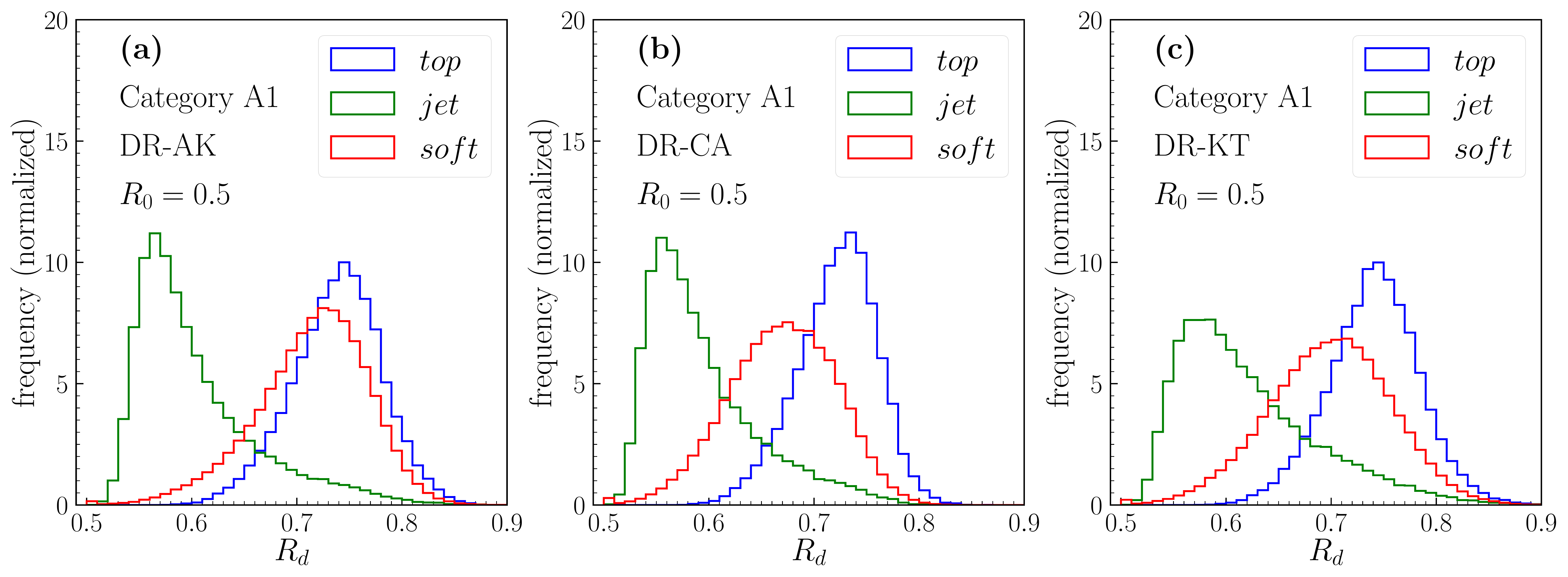}
\includegraphics[width=1.0\textwidth]{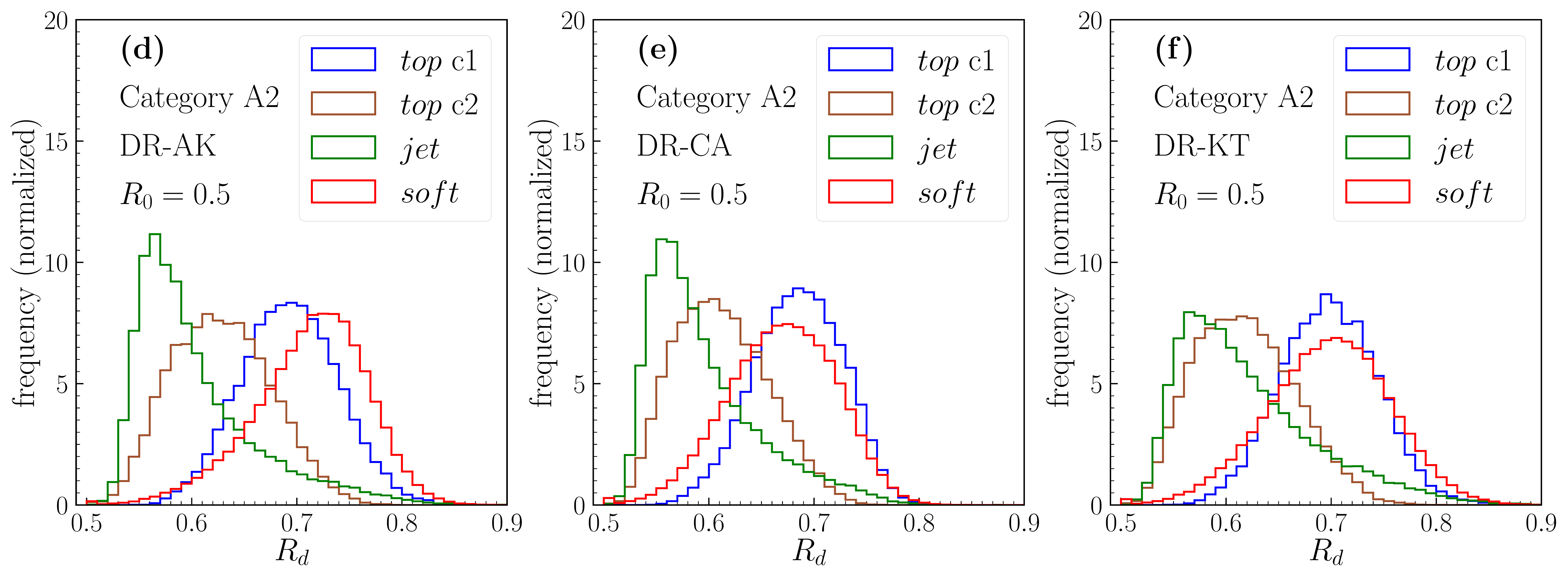}
\caption{Normalized distribution of the final radius $\Rd$ of three different
types of jets. The three plots in the top panel are for category A1 events while
those in the bottom panel are for category A2 events. The conventions for the
colours and labels `\Top', `\jet', and `\soft' are the same as in
Fig.~\ref{fig:pptj-hist}. For category A2, `\Top c1' and `\Top c2' labels
represent the two constituent jets of reconstructed \Top. The distributions are
shown for DR-AK, DR-CA, and DR-KT algorithms in the panels (a,d), (b,e), and
(e,f), respectively, with $\Rnot=0.5$.}
\label{fig:pptj-Rd-hist}
\end{figure}
%===============================================================================

An interesting feature to notice is that the mass distribution for the energetic
\jet\ remains almost the same for both the standard and dynamic radius jet
clustering algorithms. The similarity between these two are more prominent for
AK and CA algorithms and less so for the KT algorithm. This is expected as the
KT algorithm starts to merge softer momenta first and then capture the harder
ones almost at the end. As a result, this algorithm lets the size of the dynamic
radius grow in the beginning and hence allows the softer hadron, even if they
are a little wider, to merge with the evolving jet.  The \Top\ jet mass
distribution is also a little off with respect to their fixed radius analogue.
These are not very problematic since jet
grooming\,\cite{Butterworth:2008iy,Dasgupta:2013ihk,Larkoski:2014wba,Marzani:2017mva,Marzani:2017kqd,Dreyer:2018tjj,Mehtar-Tani:2019rrk},
trimming\,\cite{Krohn:2009th}, or pruning\,\cite{Ellis:2009su,Ellis:2009me}
methods help in cleaning soft and wide-angle radiation. A similar strategy of
grooming is useful in the removal of \softjets\ as well.

The change in mass distribution for \Top\ jet but not for the energetic \jet\ 
can easily be understood from the behaviour of the final radius
$\Rd=(\Rnot+\sigma)$ [Eq.~(\ref{eqn:Rdij})] a jet has acquired. We, therefore,
show the distribution of the final radii of the three different types of jets
in Fig.~\ref{fig:pptj-Rd-hist}. The three plots in the top panel are for
category A1 events while those in the bottom panel are for category A2 events.
For category A2 events, `{\it top} c1' and `{\it top} c2' labels represent the
two constituent jets of reconstructed \Top. The distributions are shown for
DR-AK, DR-CA, and DR-KT algorithms in Figs.~\ref{fig:pptj-Rd-hist}(a,d),
\ref{fig:pptj-Rd-hist}(b,e), and \ref{fig:pptj-Rd-hist}(c,f), respectively, with
$\Rnot = 0.5$ in each.

From all the histograms in Fig.~\ref{fig:pptj-Rd-hist}, some clear features
emerge. For the case of category A1 \Top\ jets, the final radius $\Rd$ grows to
more than 0.6 with a peak at $\Rd \simeq 0.75$, (approximately 50\% increase
with respect to the starting radius). On the other hand, for the energetic jets,
$\Rd$ does not grow by much. This indicates that the radius grows dynamically
according to the distribution of constituents inside the jet. The growth of the
\softjets\ is higher compared to the hard \jets\ candidates. In general, this is
will not be a problem in the heavy object finding since they can easily be
eliminated by choosing an appropriate $\pt$ or mass cuts. The story for the
category A2 events is similar for \jets\ and \softjets. The only difference is
that the whole \Top\ could not be reconstructed as a single jet in these events.
The normalized distributions of the final radii of these two constituent jets of
reconstructed \Tops\ are plotted. These constituents tend to grow more than the
energetic \jets.

%===============================================================================
%\afterpage{\clearpage}
\begin{figure}[!h]
\begin{center}
\includegraphics[width=\textwidth]{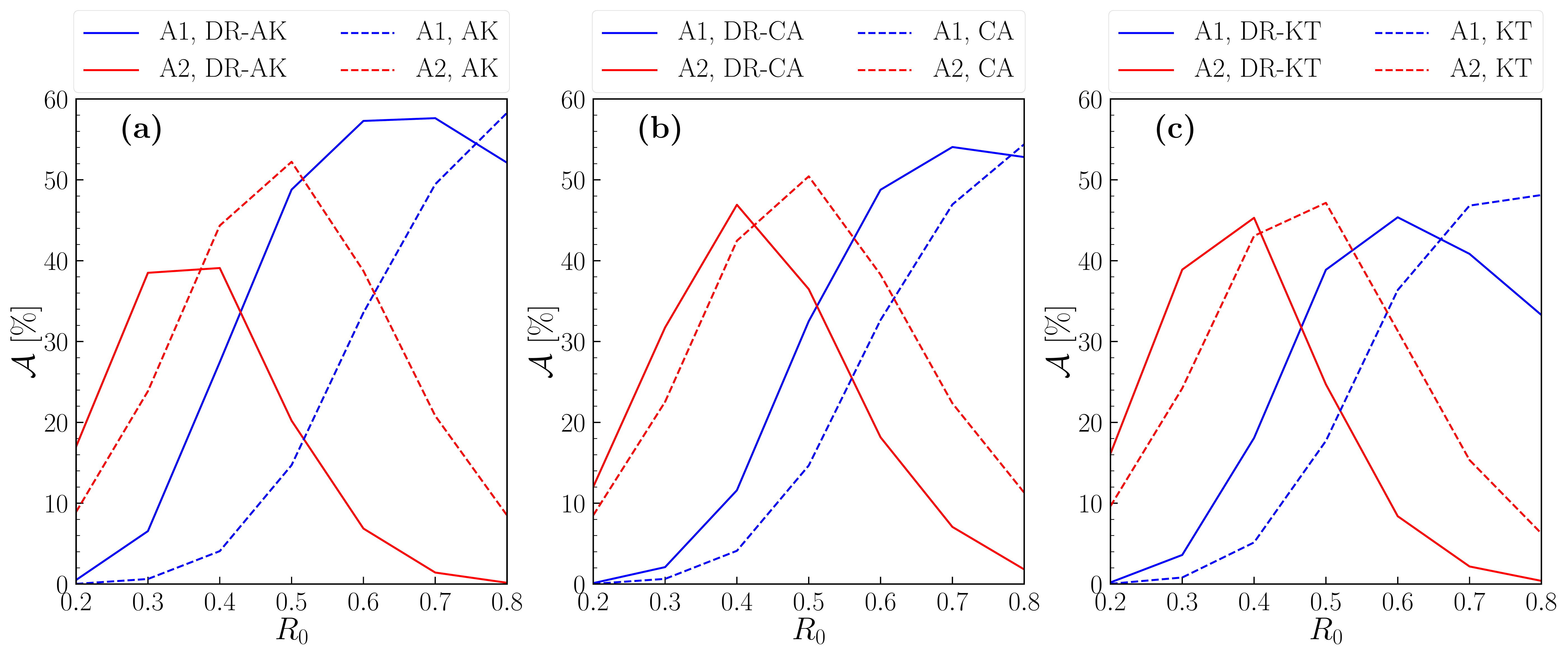}
\caption{The variation of acceptance efficiency \Acc [Eq.~(\ref{eqn:acc})] as a
function of starting radius $\Rnot$ for $pp\to tj$ SM process. The blue and red
lines represent the variations of \Acc for categories A1 and A2 events,
respectively. The dashed lines are for (a) AK, (b) CA, and (c) KT algorithm and
the solid lines are for their dynamic radius versions.}
\label{fig:tj-smmry}
\end{center}
\end{figure}
%===============================================================================

The values of acceptance efficiencies \Acc [Eq.~(\ref{eqn:acc})] for different
category events vary with the choice of the value for the starting radius
$\Rnot$. If the starting radius is small, the algorithms fail to capture the
fat jet. On the other hand, the large starting radius $\Rnot$ will capture the
unwanted contamination coming from underlying events or radiations from other
nearby showers. As a result, the jets will be unnecessarily fat and massive.
There is a suitable range for $\Rnot$ within which the algorithms work better.
We, therefore, show the variation of acceptance efficiencies \Acc as a function
of starting radius $\Rnot$ in Fig.~\ref{fig:tj-smmry} for both categories A1
(blue) and A2 (red). The variations are shown for (DR-) AK, CA, and KT
algorithms in panels (a), (b), and (c), respectively. As expected, for small
$\Rnot$ values, the efficiencies for category A1 (blue lines) are negligible in
both dynamic radius and fixed radius analyses since the constituents of the
entire \Top\ jet could not be captured with these small values of $\Rnot$.
Rather, the category A2 (red lines) which form the \Top\ with the help of two
jets yields more \Acc. This picture changes once we tend towards higher values
for $\Rnot \simeq 0.5$ as more and more \Top\ jets are being reconstructed in
the A1 category. As a result, the values of \Acc for the A2 category get
reduced. In all the panels of Fig.~\ref{fig:tj-smmry}, it is interesting to note
that the dynamic radius algorithms (solid) yield higher values for \Acc than
their fixed radius counterparts (dashed). This is indicative of the usefulness
of the dynamic radius algorithm over the fixed radius ones. The dip in the blue
solid lines after near $\Rnot = 0.7$ is not essentially the failure of the
algorithm. Rather, it is because of the capture of unwanted contaminations along
with the radiation coming from the top. Therefore, the jet mass goes beyond
200~GeV, at which point we stop labelling them as a reconstructed \Top\ jet.
Furthermore, a rough comparison among the curves in the three panels
of Fig.~\ref{fig:tj-smmry} indicates that DR-AK is better suited than DR-CA and
DR-KT algorithms. 

%===============================================================================
\subsection{Illustration II: $pp\to Vj$ Subprocess}
A similar study has been performed in SM $pp\to Vj$, ($V=W$ or $Z$) processes.
In order to ensure the formation of fat jets, a lower cut of 500~GeV on the
$\pt$ of the jet has been imposed at the time of generation of parton-level
events via MG5. These events were then passed on to {\tt Pythia8} with {\it
Monash 2013 Tune}\,\cite{Skands:2014pea} tune for parton showering and
hadronization. The final state hadrons of these events were then sent to {\tt
FastJet3} for jet clustering with starting radius $\Rnot = 0.4$.

%===============================================================================
\afterpage{\clearpage}
\begin{figure}[!h]
\begin{center}
\begin{subfigure}{\textwidth}
\includegraphics[width=0.5\textwidth]{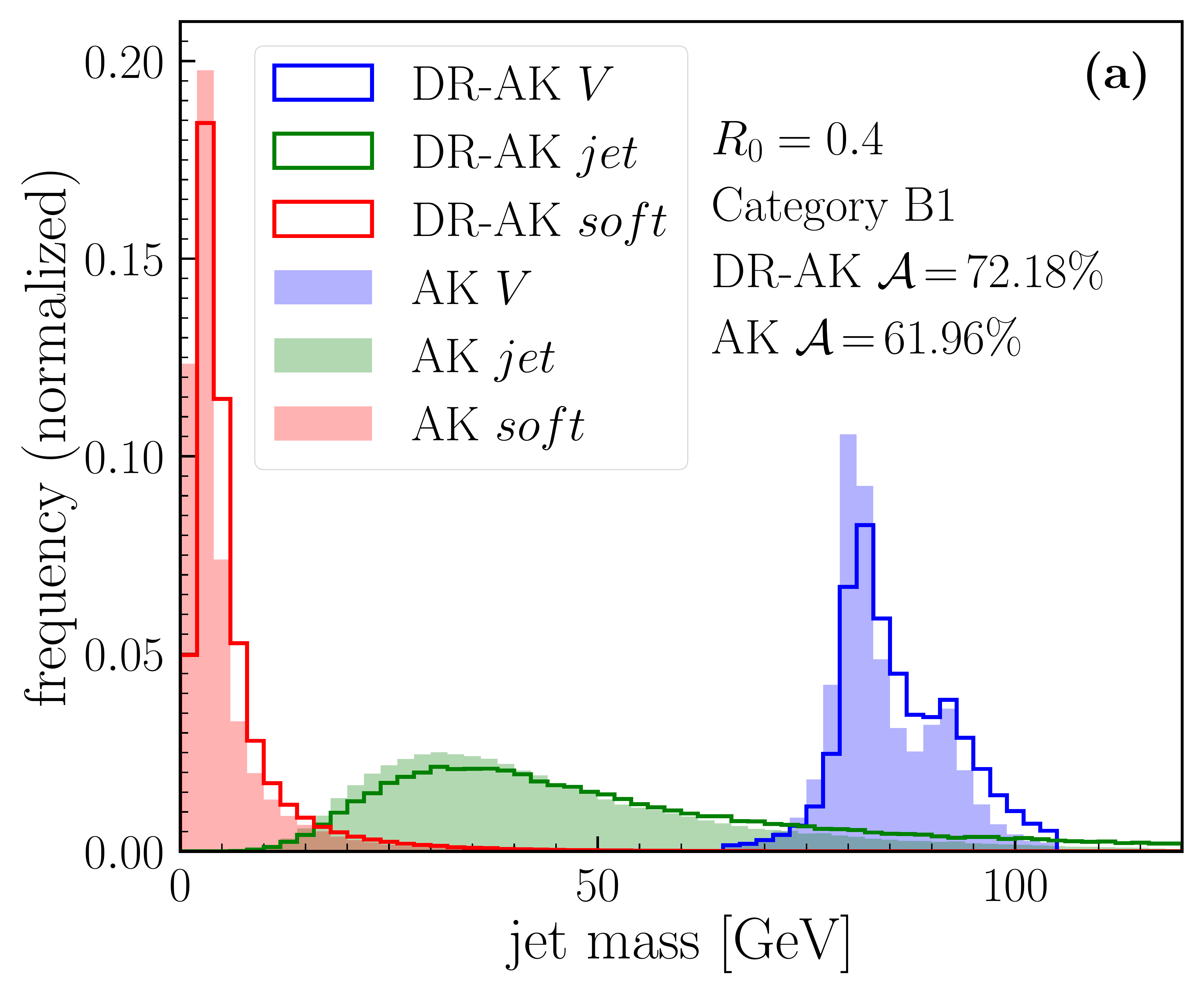}
\includegraphics[width=0.5\textwidth]{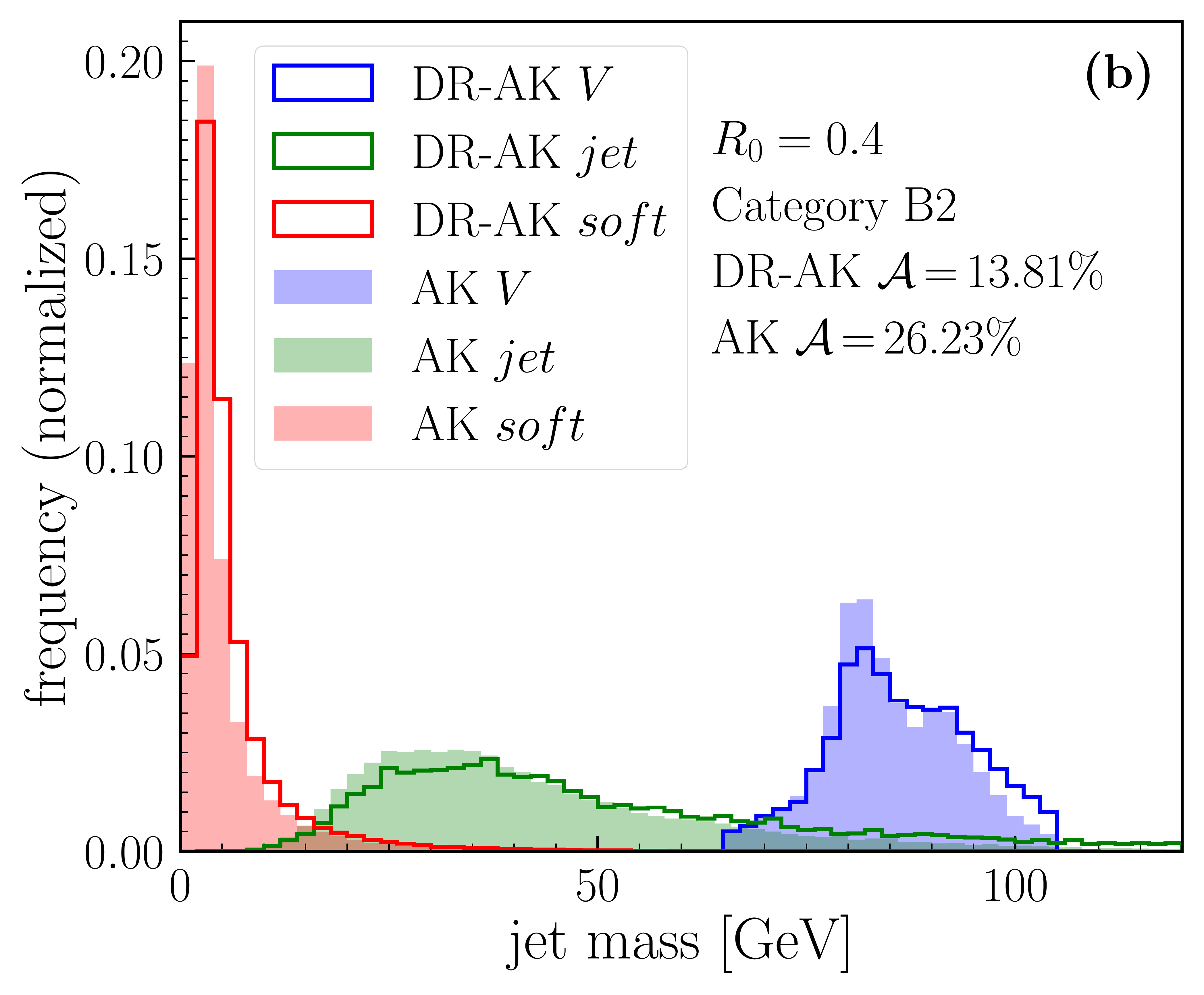}
\end{subfigure}
\begin{subfigure}{\textwidth}
\includegraphics[width=0.5\textwidth]{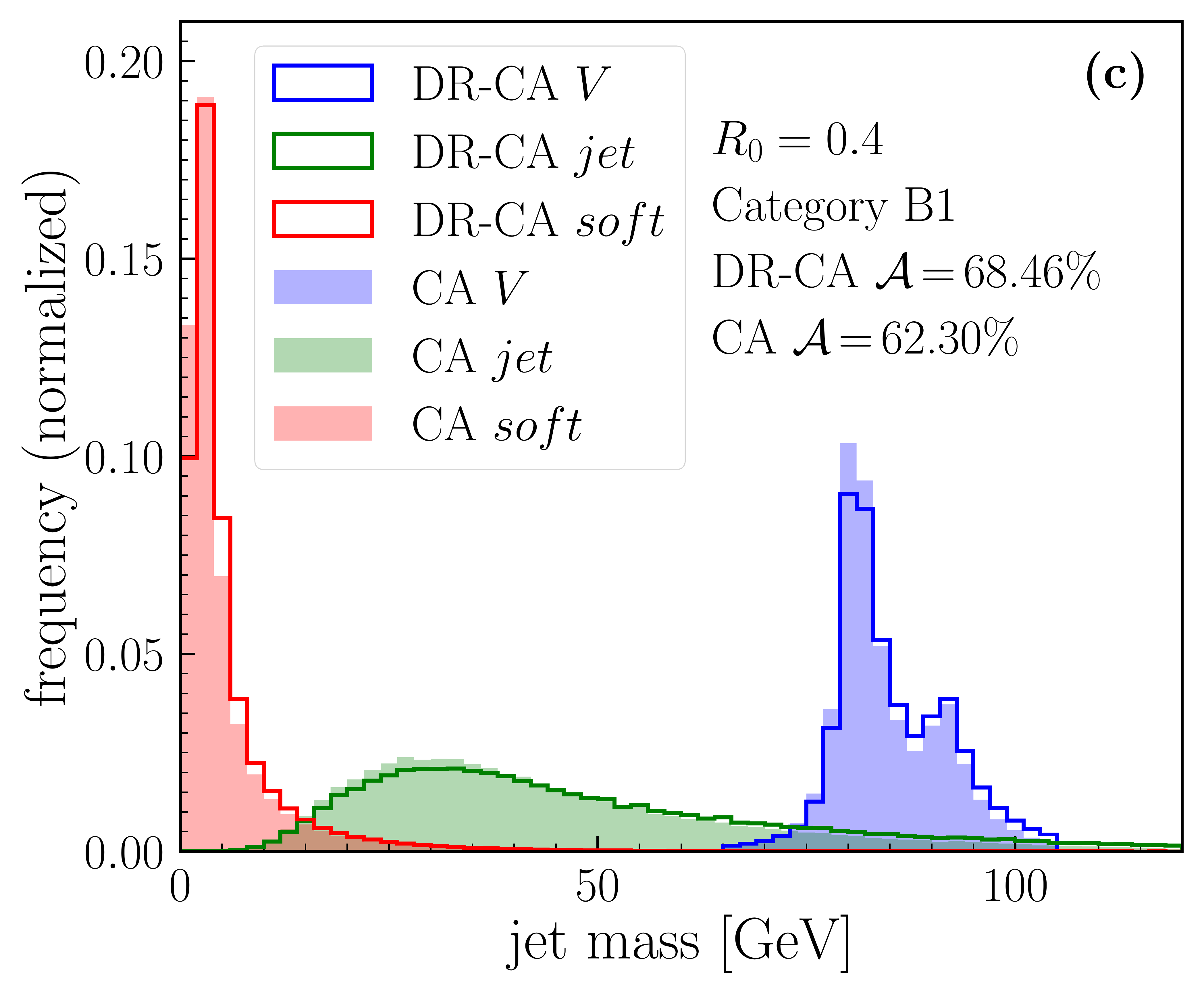}
\includegraphics[width=0.5\textwidth]{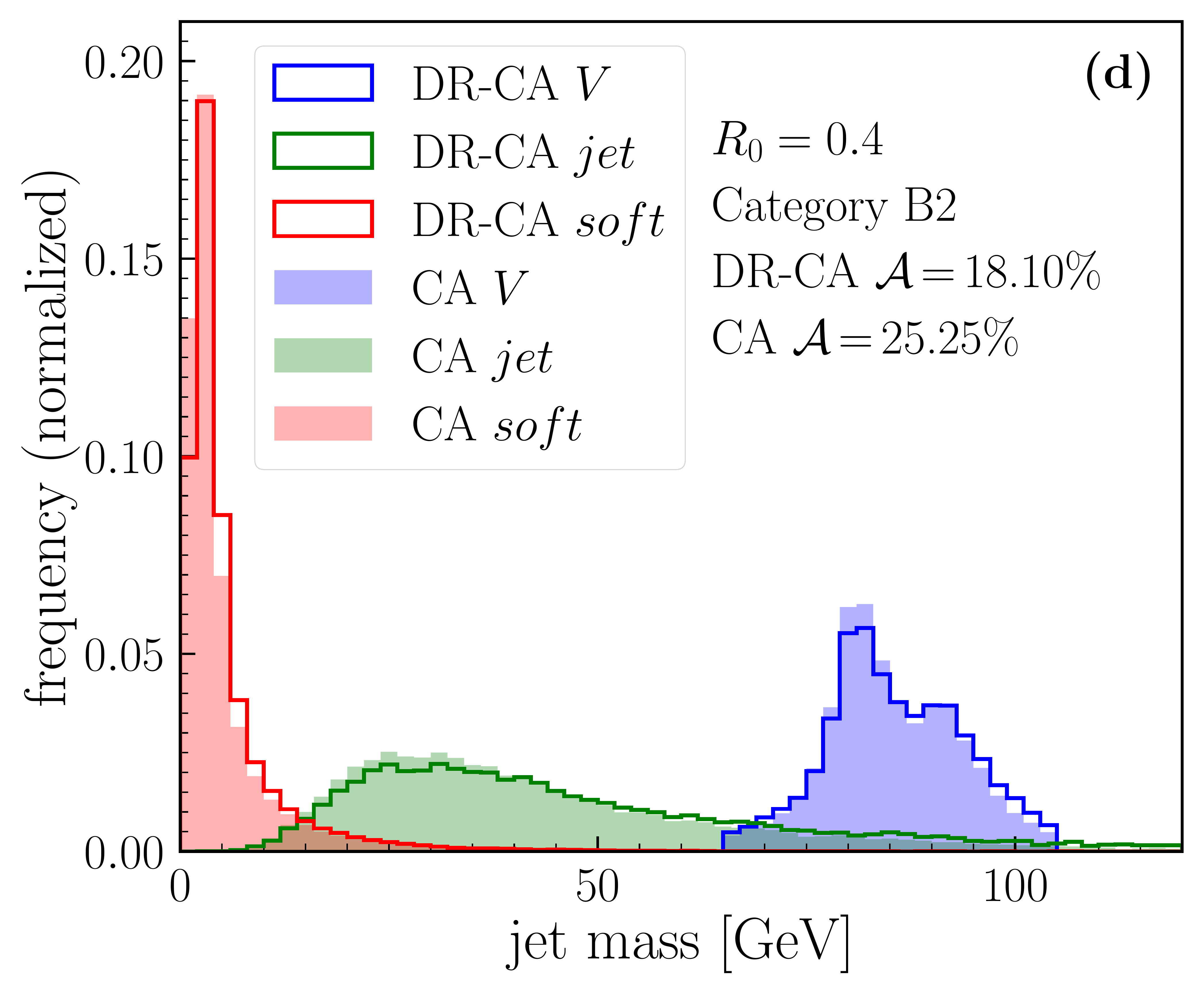}
\end{subfigure}
\begin{subfigure}{\textwidth}
\includegraphics[width=0.5\textwidth]{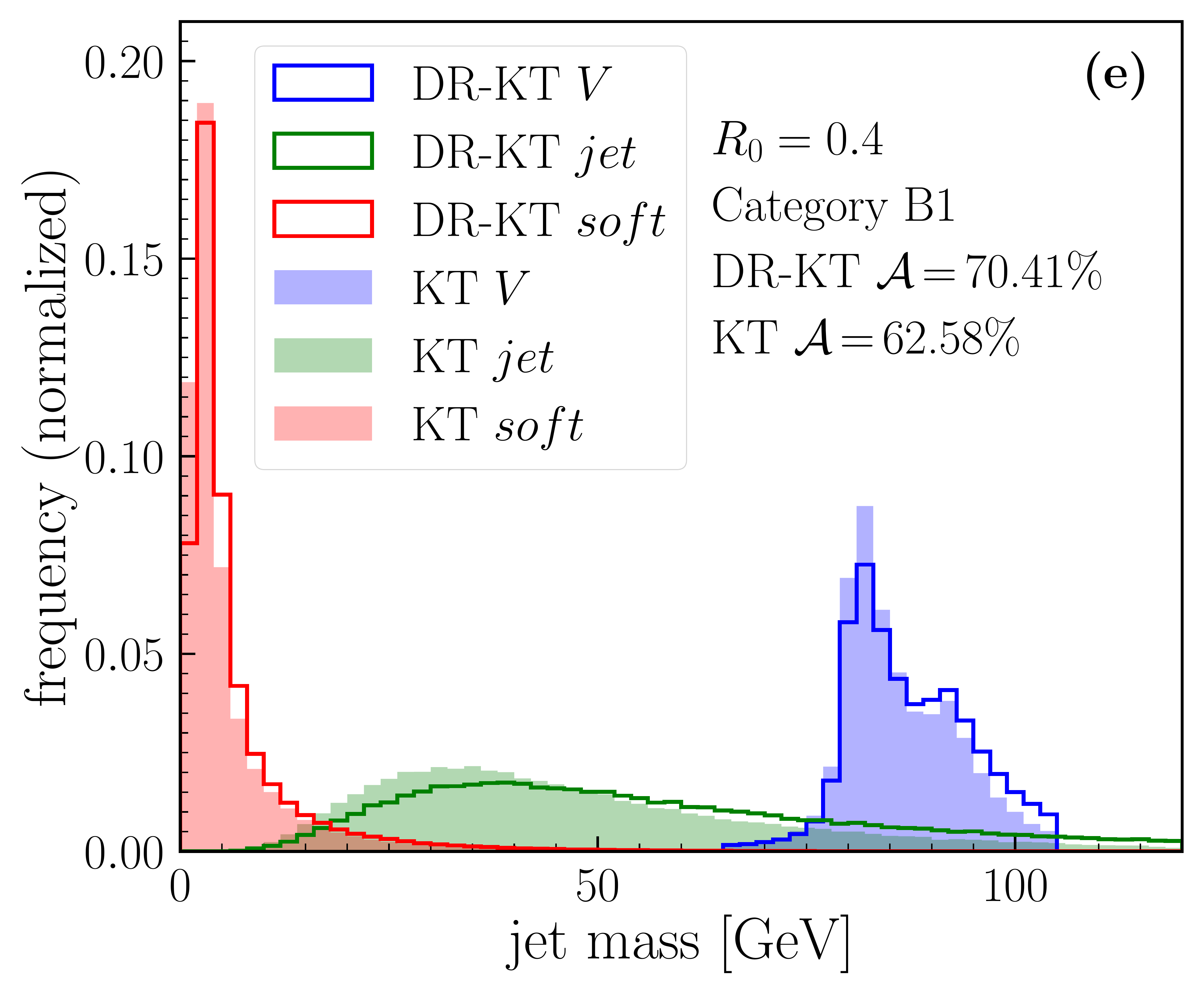}
\includegraphics[width=0.5\textwidth]{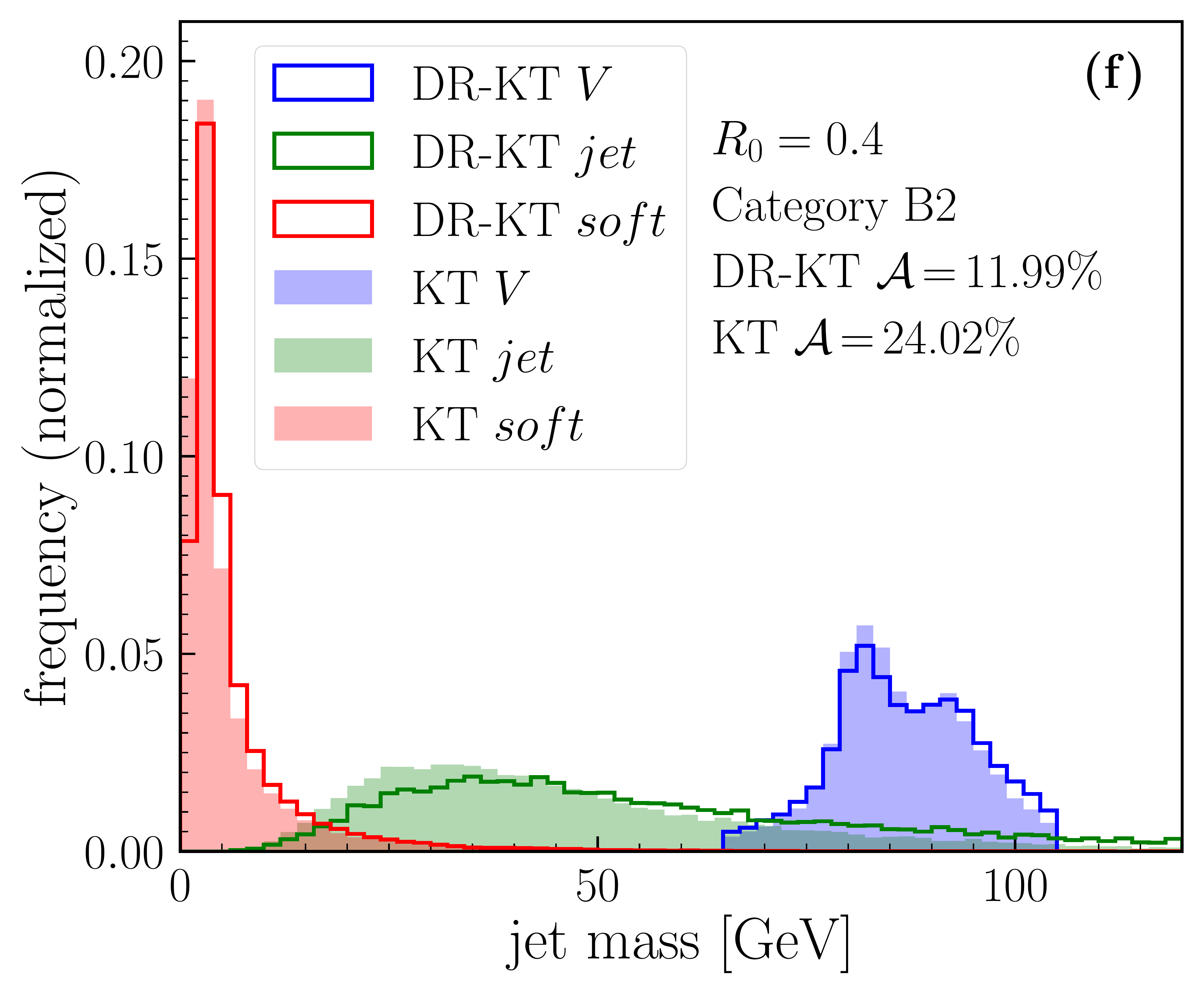}
\end{subfigure}
\end{center}%\vspace{-1cm}
\caption{Normalized distributions of jet mass for the process $pp\to V j$. The
left panel shows jet mass distributions of category B1 events and the right
panel is the distribution for category B2 events. The blue, green, and red
histograms are for reconstructed \V, energetic \jet\ and \softjets\ (defined in
the text), respectively. The histograms, from top to bottom, are for AK, CA, and
KT algorithms, respectively. The filled histograms correspond to fixed radius
algorithms and the unfilled ones correspond to their dynamic radius (DR)
analogues.}
\label{fig:ppvj-hist}
\end{figure}
%===============================================================================

As before, we label the energetic jets coming from a jet clustering algorithm,
as reconstructed `V' or reconstructed `\jet' with the help of \MGpartonlevel
information. The rest of the jets having $\pt > 5$~GeV are tagged as
`\softjets'. As in the previous illustration, we classify the events into two
separate categories based on the following criteria.
\begin{enumerate}
    \item [\bf B1.] An event was labelled as a category B1 event if it satisfies
    the following two conditions.
    \begin{itemize}
        \item A jet should have mass in the range (65, 105)~GeV and $\dR
        (\text{V}^\text{MG5}, \text{jet}) < 0.5$. This jet was identified as a
        reconstructed \V\ jet and we label them as `\VBone' in further
        discussions.
        \item After the tagging of the V jet, another jet should have
        $\pt>300$~GeV and $\dR (j^\text{MG5}, \text{jet}) < 0.5$. These jets are
        labelled as `\jetBone' in further discussions.
    \end{itemize}
    \item [\bf B2.] An event, after failing to satisfy the criteria for the
    category B1, could be classified as a category B2 event subject to
    satisfying the below conditions.
    \begin{itemize}
        \item Two separate jets within 1.0 distance from the original vector
        boson ($W$ or $Z$) and should have an invariant mass between 65 and
        105~GeV. These two jets are tagged as constituent jets of the
        reconstructed `V' jet. The final reconstructed `V' jet should be within
        0.5 distance from the original boson. This combination is labelled as
        `\VBtwo'.
        \item Another jet having $p_T > 300$~GeV and within 0.5 radii of the
        original jet and this is labelled as `\jetBtwo'.
	\end{itemize}
\end{enumerate}

We show the jet mass distribution in Fig.~\ref{fig:ppvj-hist} for SM $pp\to V j$
process. All the distributions in the left panel of the figure represent the
category B1 events and the distributions in the right panel are for category B2.
The blue, green, and red histograms are for reconstructed `V', \jet\, and
\softjets, respectively. The histograms, from top to bottom, are for AK, CA, and
KT algorithms, respectively. The filled histograms are for standard jet
clustering algorithms and the unfilled ones are their dynamic radius analogues.
Quite clearly, the two peaks in the blue histograms, in all the distributions,
correspond to the mass peaks of $W$ and $Z$ bosons. The jet mass distribution of
the energetic jets using dynamic radius algorithms remains similar to their
fixed radius counterparts. The increment in the percentage of the acceptance
efficiencies \Acc [Eq.~(\ref{eqn:acc})] of category B1 events is representative of
the appropriateness of using the dynamic radius algorithms over the standard
ones in these types of scenarios.

%===============================================================================
%\afterpage{\clearpage}
\begin{figure}[!h]
\begin{center}
\includegraphics[width=1.0\textwidth]{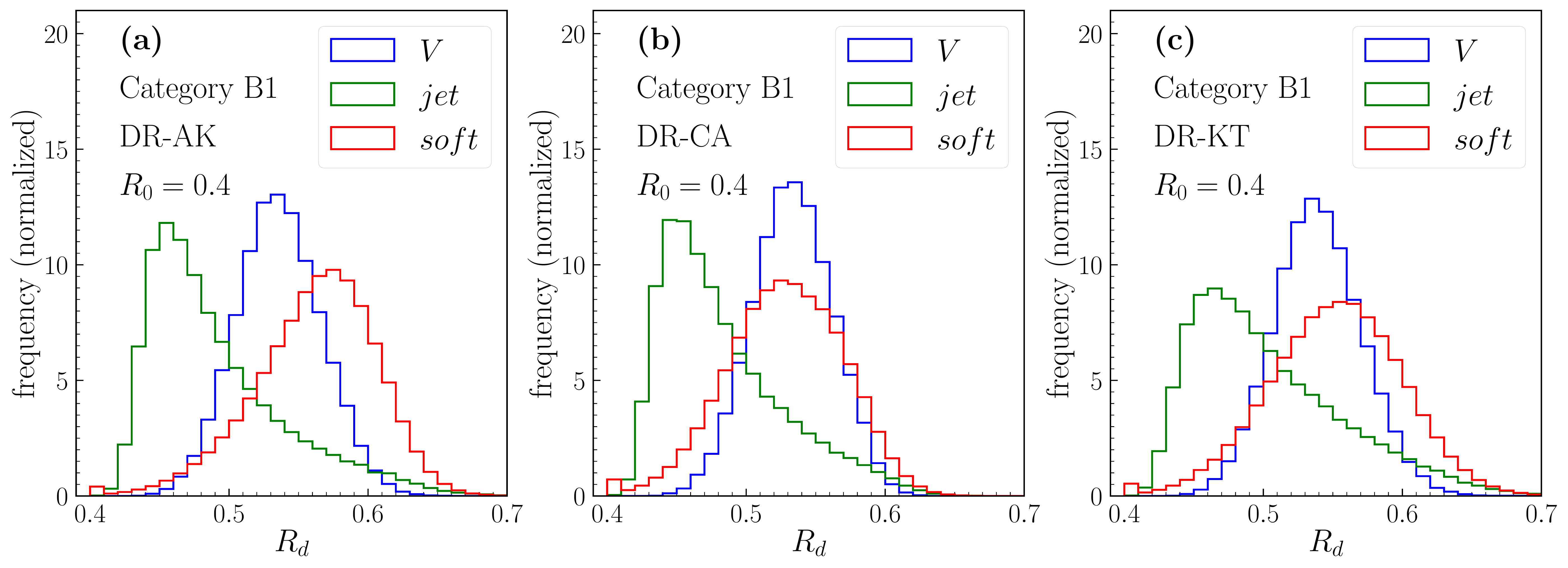}
\includegraphics[width=1.0\textwidth]{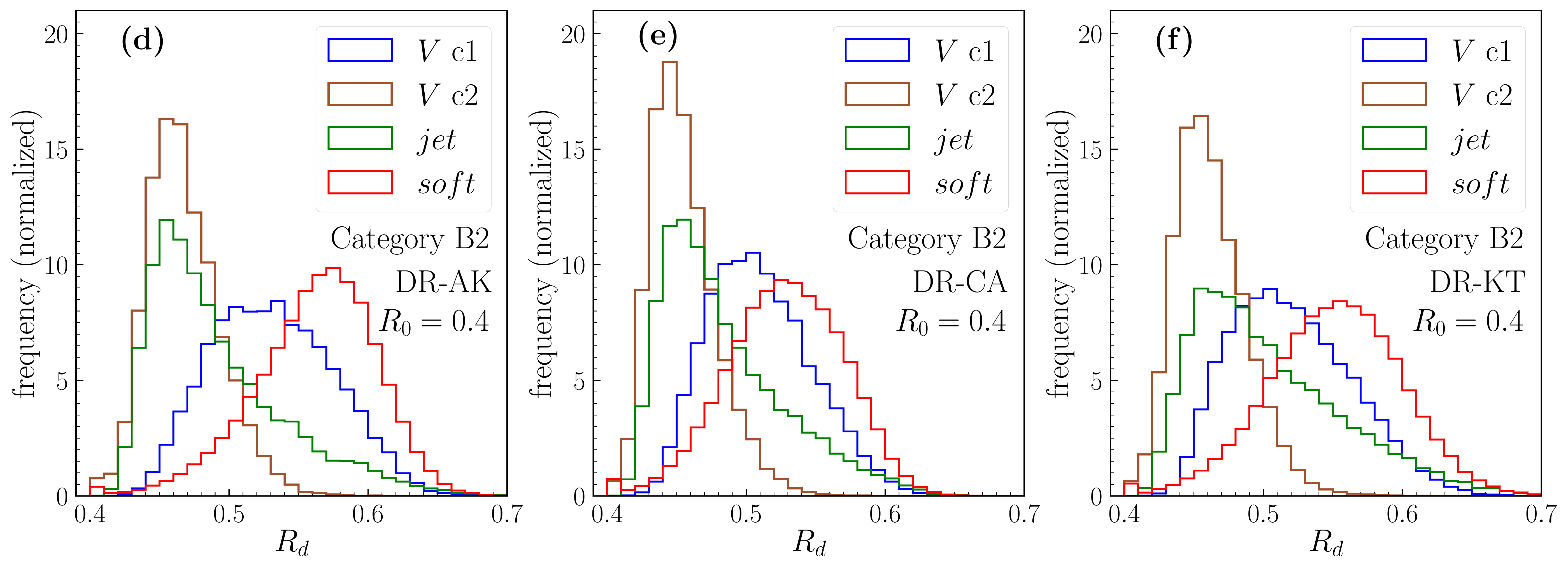}
\caption{Normalized distribution of final radius $\Rd$ for the three different
types of jets. The top panel represents the distributions of $\Rd$ in the
category B1 events and the whole bottom panel is for category B2 events. The
conventions for the colours and labels \V, \jet, and \soft\ are the same as
Fig.~\ref{fig:ppvj-hist}. For category B2 events, `\V\ c1' and `\V\ c2' labels
represent the two constituent jets of the reconstructed vector bosons. The
distributions are shown for DR-AK, DR-CA, and DR-KT algorithms in the panels
(a,d), (b,e), and (c,f), respectively, with $\Rnot = 0.4$.}
\label{fig:ppvj-Rd-hist}
\end{center}
\end{figure}
%===============================================================================

We next show in Fig.~\ref{fig:ppvj-Rd-hist} the normalized distributions of the
final radius for three different types of jets, {\it viz.} `V' jets, energetic
\jets, and \softjets. As in the $pp \to t j$ process, the fat \V\ jets acquires
a larger radius than the energetic \jets\ after the dynamical expansion of the
jet size. Here, again, the \softjets\ acquire a higher radius compared to the
energetic \jets. These \softjets are not of much concern since they are rather
soft and hence they can be removed easily from the analysis.

\begin{figure}[!h]
\begin{center}
\includegraphics[width=\textwidth]{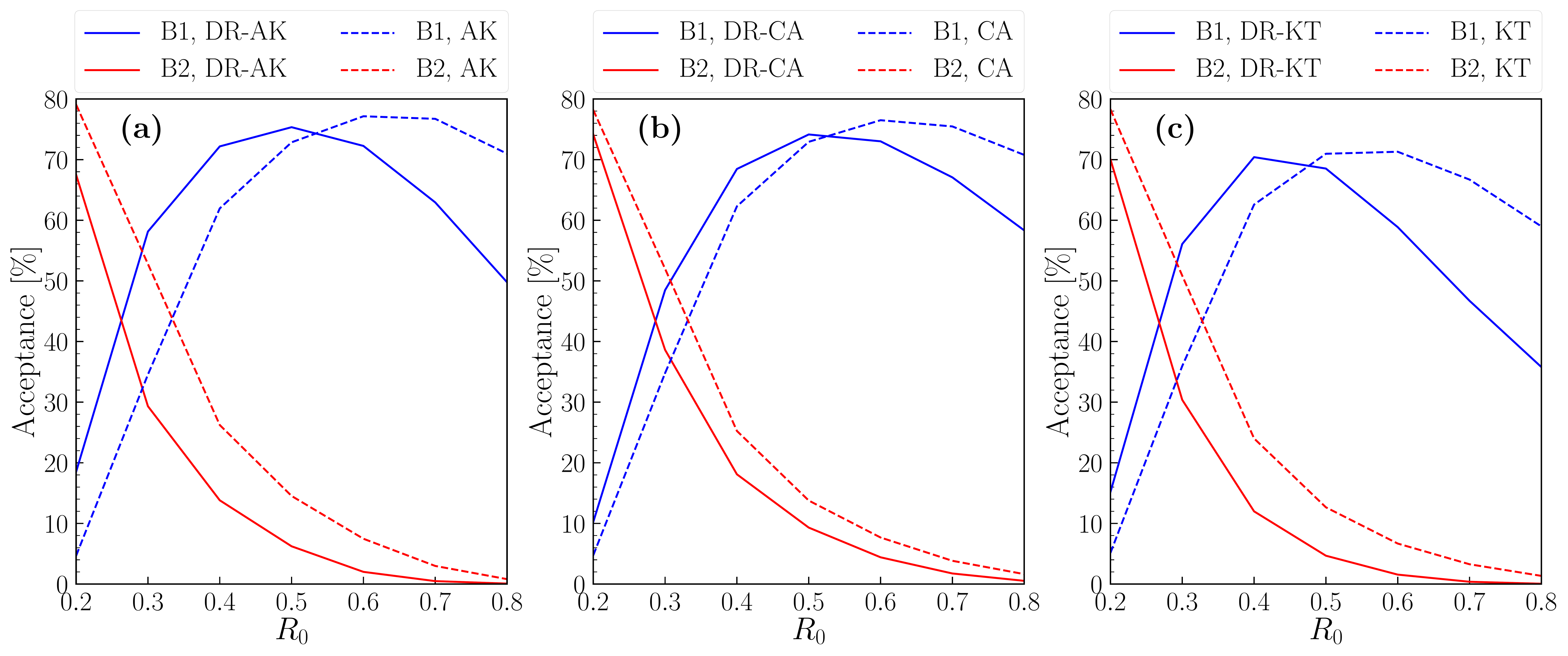}
\caption{The variation of \Acc [Eq.~(\ref{eqn:acc})] as a function of the
starting radius $\Rnot$ for $pp\to Vj$ SM process. The blue and red lines
represent the values of \Acc for categories B1 and B2 events, respectively. The
dashed lines are for (a) AK, (b) CA, and (c) KT algorithms. The solid lines are
for their dynamic radius versions.}
\label{fig:vj-smmry}
\end{center}
\end{figure}

In Fig.~\ref{fig:vj-smmry}, we show the variation of \Acc [Eq.~(\ref{eqn:acc})]
as a function of starting radius $\Rnot$. In all the panels of the figure, the
blue and red lines correspond to the variations for categories B1 and B2 events,
respectively. The dashed lines are for fixed radius algorithms and the solid
lines are for dynamic radius jet algorithms. The variations are shown for (a)
AK, (b) CA, and (c) KT algorithms. A quick observation of the curves tells us
that the behaviour of these curves is similar to that of the curves in
Fig.~\ref{fig:tj-smmry} except the monotonic decreasing nature of the category
B2 curves. The reason is as follows: in the case of \V\ jets, the jets are
`two-pronged' in nature. Therefore, the small radius jets can capture one of the
two prongs of \V\ jets, and thereby these two jets are able to reconstruct \V\
jets in B2 category. However, as the starting radius $\Rnot$ is increasing, more
and more events are migrating to category B1. The declining nature of the curves
for large radii after 0.5 is because of the fact that the jets capture
more hadrons than are required for their optimal size. As a result, the mass of
the \V\ jets tends to go beyond the mass window set to label them as \V\ jets.
Again, more variables than just the jet mass can help one to improve the tagger
and hence the acceptance efficiency.

%===============================================================================
%===============================================================================

We conclude this section with the note that the dynamic radius jet algorithms
are useful in finding fat as well as narrow jet in a single event in the
colliders. We have successfully illustrated this in two SM processes, {\it viz.}
$pp\to tj$ and $pp\to Vj$, at the  13~TeV LHC. A comparison among the three
dynamic radius analogues of the standard $\kt$-type algorithm reveals that the
DR-AK algorithm performs better compared to the DR-CA or the DR-KT algorithms.

%===============================================================================
\section{Usefulness in BSM signals}\label{sec:appBSM}
We now illustrate the usefulness of the dynamic radius jet algorithm in the
context of a scenario beyond the standard model (BSM). This is a scenario where
an additional vectorlike singlet quark $b'$ of charge $-1/3$ exists along with
$(d ,s, b)$. Such quarks occur, for example, in $E(6)$ grand unified theories,
as also in some seesaw models of quark
masses\,\cite{Barger:1985nq,Rizzo:1985kn,Mukhopadhyaya:1990cv,Mukhopadhyaya:1992de,Chivukula:1998wd,Gillioz:2013pba}.
The $b'$ can mix with the three SM down-type quarks when electroweak symmetry
breaking takes place\footnote{In the following discussion, we shall (a) denote
this mass eigenstate itself by $b'$, (b) assume that ordinary-exotic quark
mixing takes place involving only the third family sequential quark, namely,
$b$, and (c) parametrize the $b$-$b'$ mixing by the angle $\theta$.}. This
causes the mass eigenstate dominated by $b'$ to decay into a top quark and a $W$
boson. In addition, the mixing between a $T_3 = -1/2$ quark and one with $T_3=0$
induces flavour-changing $Z$- and Higgs-couplings in the $b$-$b'$ sector. Thus
the $b'$, produced via strong interactions at the LHC, has the decays
$b'\rightarrow tW,~b' \rightarrow bZ,~b' \rightarrow bh$. The detailed
theoretical framework and the resulting phenomenology have been discussed widely
in the
literature\,\cite{Maalampi:1986pv,Raychaudhuri:1989cy,Barger:1995dd,Andre:2003wc,Gopalakrishna:2011ef,Aguilar-Saavedra:2013qpa,Girdhar:2014wua,Ellis:2014dza,Das:2017fjf,Choudhury:2021nib}.

The currently available data from the LHC restrict $m_{b'}$ to be no less than
1.3--1.5~TeV \cite{CMS:2020ttz,CMS:2022fck,ATLAS:2022hnn,ATLAS:2022tla}. When
such a massive quark decays thereafter, its decay products are considerably
lighter compared to it. Therefore the $b'$ decay products are considerably
boosted, so as to produce fat jets. Furthermore, the difference in mass between
two product particles leads to jets of varying degrees of fatness. 

Since our purpose here is to show the efficacy of the dynamic radius jet
algorithm, we illustrate our main points in the context of $pp\rightarrow
b'\bar{b'}$ followed by each $b'$ decaying into a top quark and a $W$ boson.
The $t$'s and the $W$'s thus give rise to energetic jets of different radii. We
demonstrate below how our newly developed algorithm can capture the identity of
the ensuing final state. While the present work is aimed at capturing the
essence of our proposed jet algorithm, a more detailed discussion, including
combinations of all the three aforementioned decay channels of the $b'$, is
going to be presented in a separate work\,\cite{InPreparation}.

%===============================================================================
\begin{table}[!h]
\renewcommand{\arraystretch}{1.2}
\begin{center}
\begin{tabular}{|c|c|c|}
\hline
$m_{b'}$ & $\sin\theta_L$ & $\sin\theta_R$\\
\hline
1.3~TeV & 0.12 & $8.02\times 10^{-3}$\\
\hline
\end{tabular}
\caption{Values of some important parameters of the vectorlike singlet $b'$
model considered for the illustration.}\label{tab:param}
\end{center}
\end{table}
%===============================================================================

The model has been implemented in a Mathematica-based package {\tt
SARAH}\,\cite{Staub:2008uz,Staub:2013tta,Staub:2015kfa}. The Universal FeynRules
Output (UFO)\,\cite{Degrande:2011ua} generated by SARAH is then used in MG5
for the generation of parton-level events. The parameter card for MG5 has been
generated using spectrum generator
{\tt SPheno}\,\cite{Porod:2003um,Porod:2011nf}. The values for the important
parameters of the model are tabulated in Table~\ref{tab:param}. The angles
$\theta_L$ and $\theta_R$ in the table represent the mixing angle between SM
$b$ quark and exotic $b'$ quark of chirality left and right, respectively. After
the generation of the \MGpartonlevel events, the rest of the analysis pipeline
is the same as the previous illustrations of SM processes.

%===============================================================================
%\afterpage{\clearpage}
\begin{figure}[!h]
\begin{center}
\begin{subfigure}{\textwidth}
\includegraphics[width=0.5\textwidth]{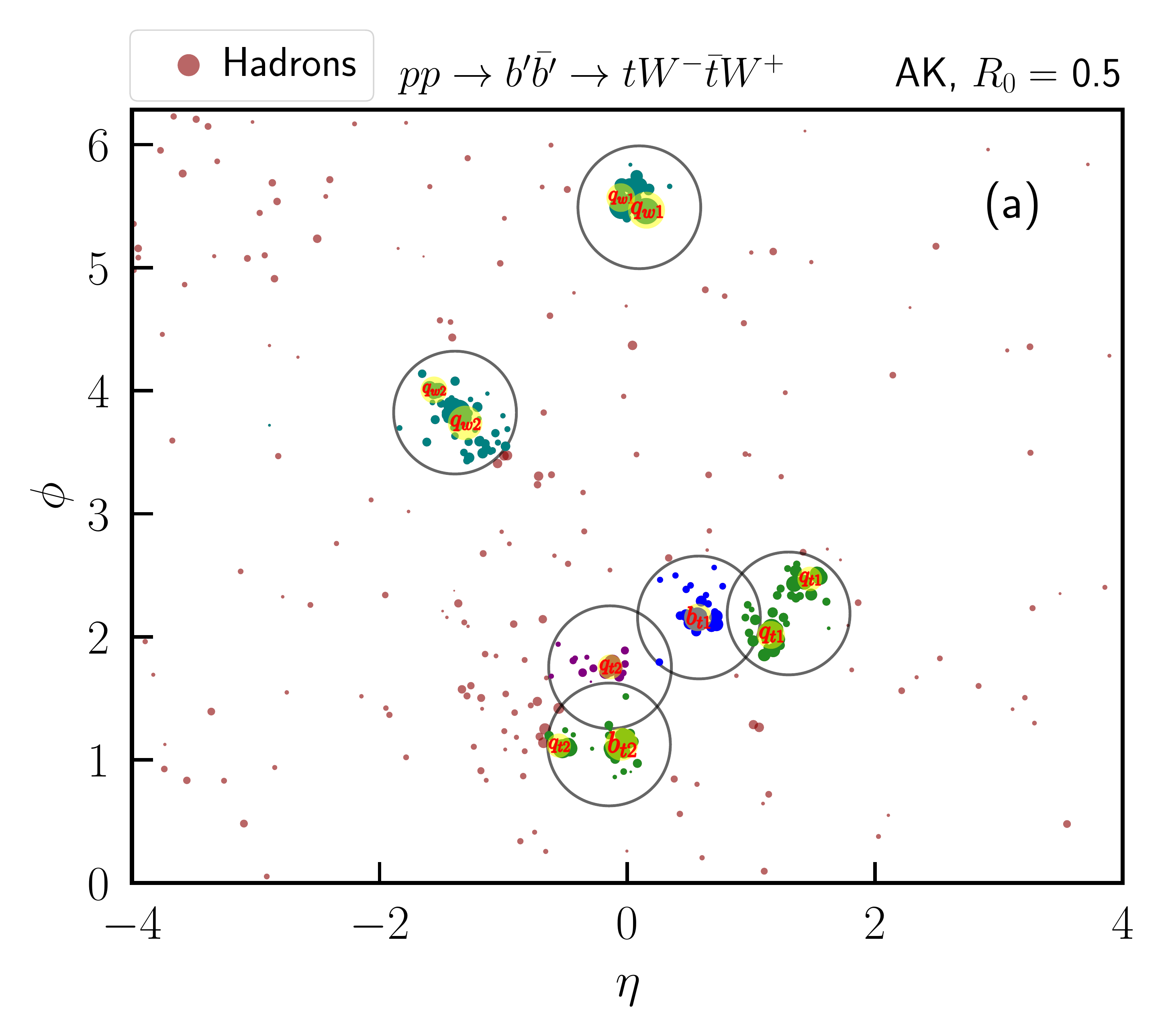}
\includegraphics[width=0.5\textwidth]{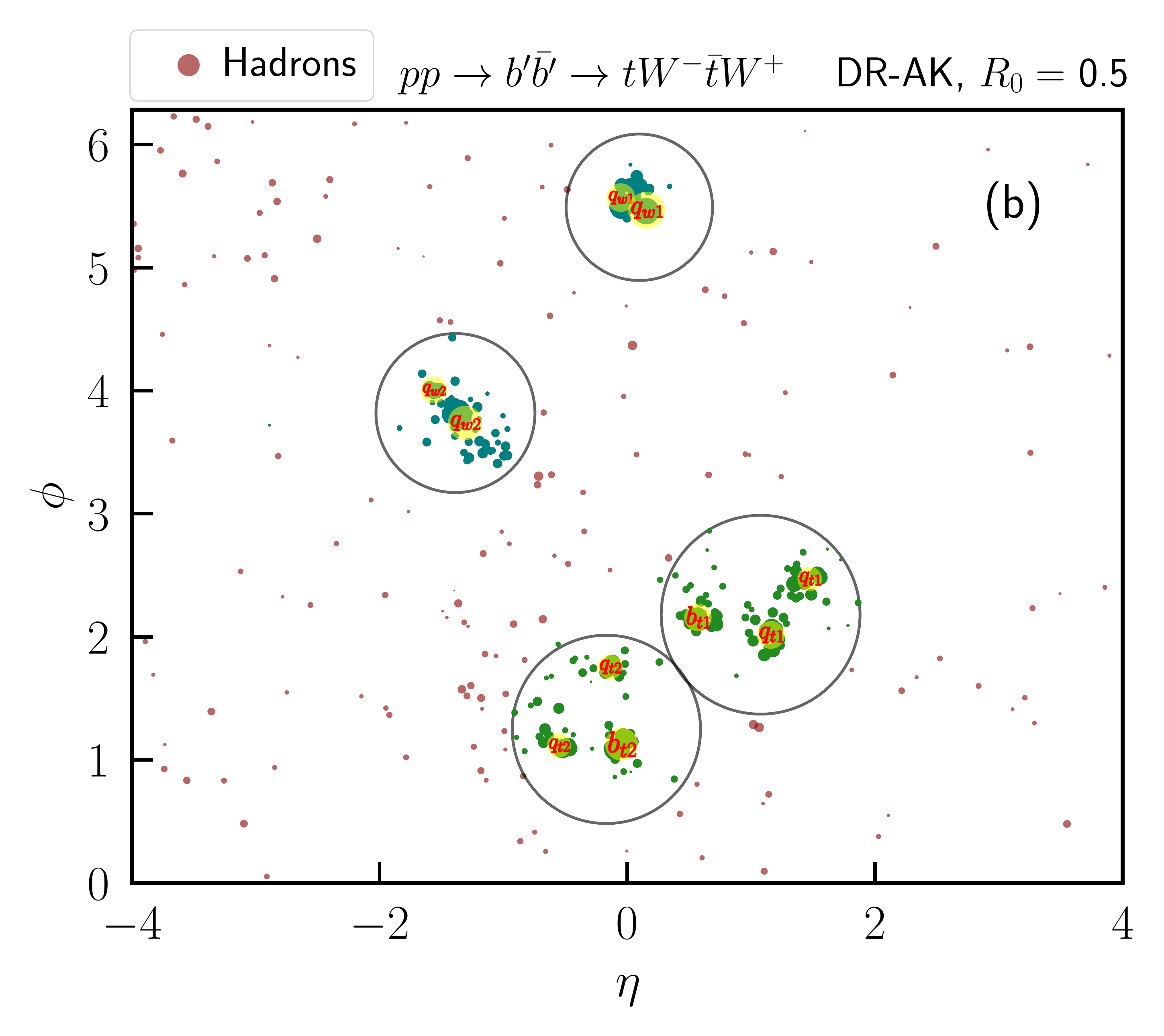}
\end{subfigure}
\caption{The distribution of final state hadrons and jets in $\eta$-$\phi$ plane
for an example event. The colours and sizes of the dots and circles follow the
same convention as Fig.~\ref{fig:tj-example}. The teal coloured dots represent
the constituents of hard fat `W' jets. The green and blue (wherever applicable)
dots represent the constituents of the fat `top' jet. The yellow dots containing
texts represent the position of the hard quarks after the decay of top or $W$
which are mentioned as the subscripts of $q$ or $b$. The plots are shown for (a)
AK and (b) DR-AK algorithms.}
\label{fig:twtw-example}
\end{center}
\end{figure}
%===============================================================================

In this illustration, we choose DR-AK, based on the discussion in the previous
section. We show the resultant jets having $\pt > 30$~GeV formed out of the
hadrons generated by Pythia8 in Fig.~\ref{fig:twtw-example}. The left panel
shows the positions of the generated hadrons and jets constructed using the AK
algorithm with $\Rnot = 0.5$. The right panel shows the same for the DR-AK
algorithm. In both panels, the red dots represent the position of final state
hadrons in the $\eta$-$\phi$ plane and the size of each dot is proportional to
the $\sqrt{\pt}$ of the hadron. The unfilled circles represent the final radius
($\Rd$) of a jet. The teal dots represent the constituents of boosted
fat `W' jets. The green, blue, and purple (wherever applicable) dots represent
the constituents of the fat `top' jet. The yellow dots containing
texts represent the position of the \MGpartonlevel $\pt$-hard quarks after the
decay of top or $W$. The mothers of the $q$ or $b$ are mentioned in the
subscripts of $q$ or $b$.

An interesting point to observe in Fig.~\ref{fig:twtw-example}(b) is that the
DR-AK yields only 4 jets, which are representative of 2 fat $W$ and 2 fat $t$
jets. However, in Fig.~\ref{fig:twtw-example}, the fixed radius algorithm could
form the fat $W$ jets but fails to capture the entirety of the two fat $t$ jets.
One, of course, can use a bigger radius in the AK algorithm to capture the whole
of the top jet. However, this will make the $W$ jet unnecessarily fat. This
demonstrates the utility of the dynamic radius jet algorithm. 

To study the goodness of DR-AK quantitatively, we define the following criteria
for tagging of \Top\ and $W$ jets.
\begin{itemize}
    \item A jet having mass in the range (150, 200)~GeV and having
    $\dR(\text{top}^\text{truth}, \text{jet}) < 0.5$ is identified as a
    reconstructed \Top\ jet.
    \item A jet will be called \W\ jet if it has a mass in the range (65, 105)~GeV
    and is within 0.5 distance from the original \MGpartonlevel $W$ boson.
\end{itemize}

Similar to the illustrations with SM processes, we classify the events into
different categories. Due to the complex nature of the final states, we have
classified the events into more than two categories in the present scenario. The
realization is based on the following understanding.
\begin{itemize}
    \item Out of the two $W$'s coming directly from $b'$ in an event, the number
    of reconstructed $W$ as fat jet from the algorithm could be 0, 1, or 2. We
    call these reconstructed fat \W\ jets as primary \W\ jets.
    \item Similarly, out of the two $t$ quarks, the number of reconstructed $t$
    as fat jets can be 0, 1, or 2. 
    \item In some particular cases, the whole top may not be reconstructed, but
    the $W$ boson coming from the top quarks may be reconstructed. These are
    referred to as secondary \W\ jets in the subsequent discussions.  
\end{itemize}

Based on the above observations, we classify the events into different
categories, whose generic name is given as C$ij$, where $i$ and $j$ are two
integers encoding the number of reconstructed \Top\ and reconstructed \W's,
respectively. For the present scenario, the allowed value for $i$ does not
exceed two. For a given $i$, the values for $j$ should not exceed $4-i$. That
is, to say, $i \le 2$ and $j \le 4-i$. An exhaustive list of all possible
categories is tabulated in Table~\ref{tab:twtw-category}. For example, the
event shown in Fig.~\ref{fig:twtw-example} would be categorized as C22 for the DR-AK
algorithm while the same event would be classified as C03 for the AK algorithm. One
may again subdivide some of the categories into subcategories based on how many
\W\ jets are coming directly from $b'$ (primary \W) and how many of them are
coming from the decay of the top quark (secondary \W). Therefore, the generic
name for the subcategories can be given as C$ijk$ with $i$, $j$, and $k$ being
the numbers of reconstructed \Top, primary \W, and secondary \W\ jets. The
possible ranges for $i, j,$ and $k$ are $0\le i,j \le 2$ and $0\le k \le 2-i$.

%===============================================================================
\begin{table}[!h] \renewcommand{\arraystretch}{1.3}
\begin{center}
\begin{tabular}{|c|c|c|c|c|}
\hline
\multirow{2}{*}{Category} & \multirow{2}{*}{Subcategory} & \multirow{2}{*}{No.~of \Top\ jet} & No. of primary & No.~of secondary \\
&&& \W\ jet &  \W\ jet \\
\hline
C22 & C220 & 2 & 2 & 0 \\
\hline
C21 & C210 & 2 & 1 & 0 \\
\hline
C20 & C200 & 2 & 0 & 0 \\
\hline
C13 & C121 & 1 & 2 & 1 \\
\hline
\multirow{2}{*}{C12} & C120 & 1 & 2 & 0 \\
\cline{2-5}
& C111 & 1 & 1 & 1\\
\hline
\multirow{2}{*}{C11} & C110 & 1 & 1 & 0\\
\cline{2-5}
& C101 & 1 & 0 & 1\\
\hline
C10 & C100 & 1 & 0 & 0\\
\hline
C04 & C022 & 0 & 2 & 2 \\
\hline
\multirow{2}{*}{C03} & C021 & 0 & 2 & 1\\
\cline{2-5}
& C012 & 0 & 1 & 2\\
\hline
\multirow{3}{*}{C02} & C020 & 0 & 2 & 0 \\
\cline{2-5}
& C011 & 0 & 1 & 1\\
\cline{2-5}
& C002 & 0 & 0 & 2\\
\hline
\multirow{2}{*}{C01} & C010 & 0 & 1 & 0 \\
\cline{2-5}
&C001 & 0 & 0 & 1\\
\hline
C00 & C000 & 0 & 0 & 0\\
\hline
\end{tabular}
\caption{The definitions of the list of categories and subcategories as
according to how many fat jets can be reconstructed from the jet algorithm.}
\label{tab:twtw-category}
\end{center}
\end{table}
%===============================================================================

%===============================================================================
%\afterpage{\clearpage}
\begin{figure}[!h]
\begin{center}
\includegraphics[width=\textwidth]{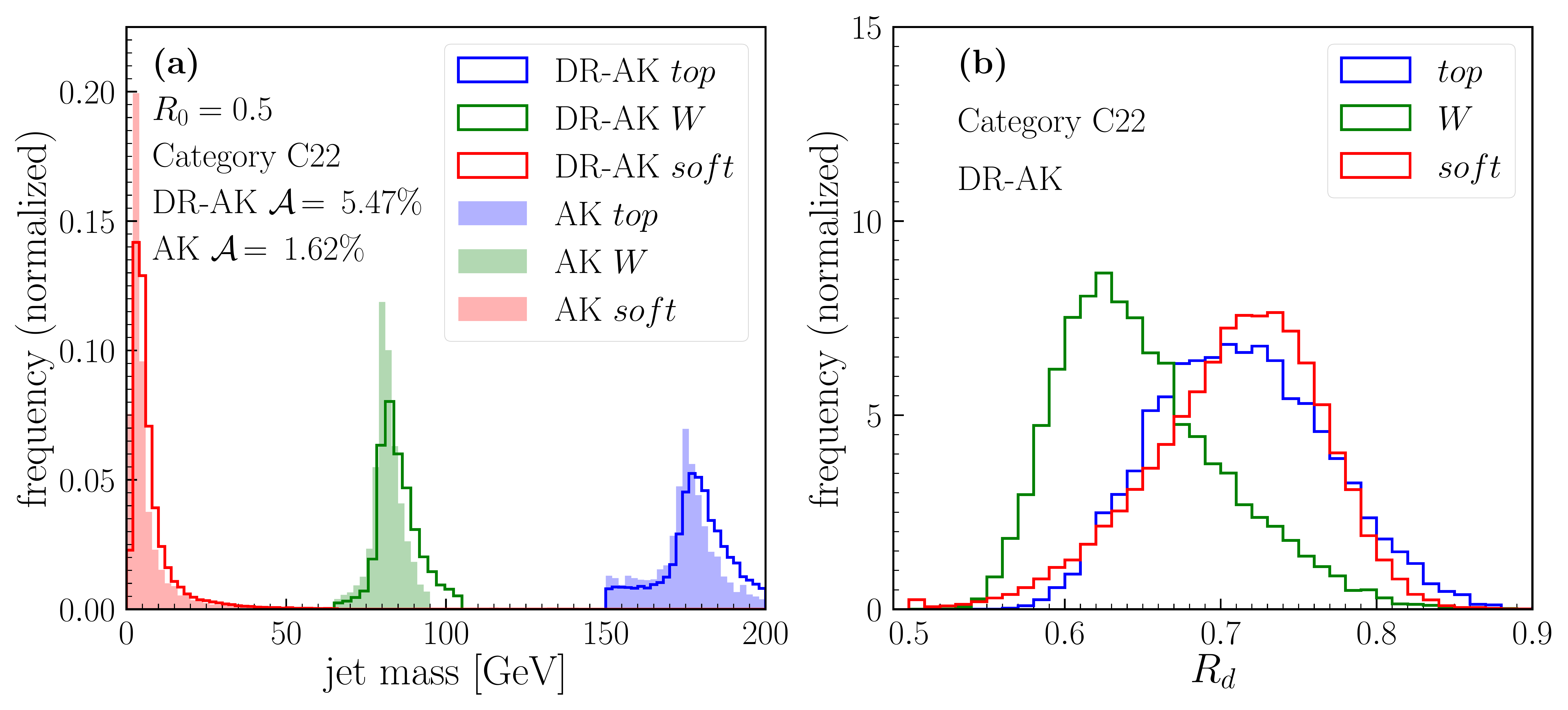}
\caption{(a) The normalized distribution of jet mass of the category C22 events
for the $pp\to b'\bar{b'} \to tW^-\bar tW^+$ process. The blue, green, and red
histograms are reconstructed \Top, \W, and \softjets, respectively. The unfilled
histograms are for the jets clustered using the DR-AK algorithm while the filled
ones are for the jets using the AK clustering algorithm. (b)~The normalized
distribution of the final radii of \Top, \W, and \softjets\ with blue, green,
and red colours, respectively. For both panels, $\Rnot=0.5$ was used and any
additional jets having $\pt\! >\! 5$~GeV were considered as a \softjet.}
\label{fig:twtw-mJ-Rd}
\end{center}\vspace{-5mm}
\end{figure}
%===============================================================================

We plot the normalized distribution of jet mass of the category C22 events in
Fig.~\ref{fig:twtw-mJ-Rd}(a). Jets were clustered using $\Rnot=0.5$. In the
plot, the blue, green, and red histograms are reconstructed \Top, \W, and
\softjets, respectively. The unfilled histograms are for the jets clustered
using the DR-AK algorithm, and the filled ones are for the jets using the AK clustering
algorithm. Any untagged jet with $\pt\!>\!5$~GeV was considered to be a
\softjet. Fig.~\ref{fig:twtw-mJ-Rd}(b) shows the normalized distribution for
the finally acquired radii of different jets for category C22 events. The
desirable feature of the reconstructed \W\ jets being narrower than the
reconstructed \Top\ jets is clearly apparent in the figure. Here, again, the
\softjets\ are growing to larger radius are expected. However, as discussed in
the previous section, they can be removed from an analysis by $\pt$ or jet mass
cuts. 

The values of \Acc [Eq.~(\ref{eqn:acc})] for the two different algorithms, {\it
viz.} DR-AK and AK are also quoted in Fig.~\ref{fig:twtw-mJ-Rd}(a). These values
(1.62\% for AK and 5.47\% for DR-AK), clearly, indicate that the dynamic radius
jet algorithm is working better while probing the correct mass windows for the
particles. The shift of the mass distribution towards larger values is
indicative of capturing little extra than required. As discussed previously,
this can be rectified by the techniques of
grooming\,\cite{Larkoski:2014wba,Butterworth:2008iy,Dasgupta:2013ihk,Marzani:2017mva,Marzani:2017kqd},
trimming\,\cite{Krohn:2009th}, or pruning\,\cite{Ellis:2009su,Ellis:2009me}.
Furthermore, going beyond just the jet mass to tag the \Top or $W$ jets would
further help in extracting signals.

%===============================================================================
\begin{figure}[!h]
\begin{center}
\includegraphics[width=0.98\textwidth]{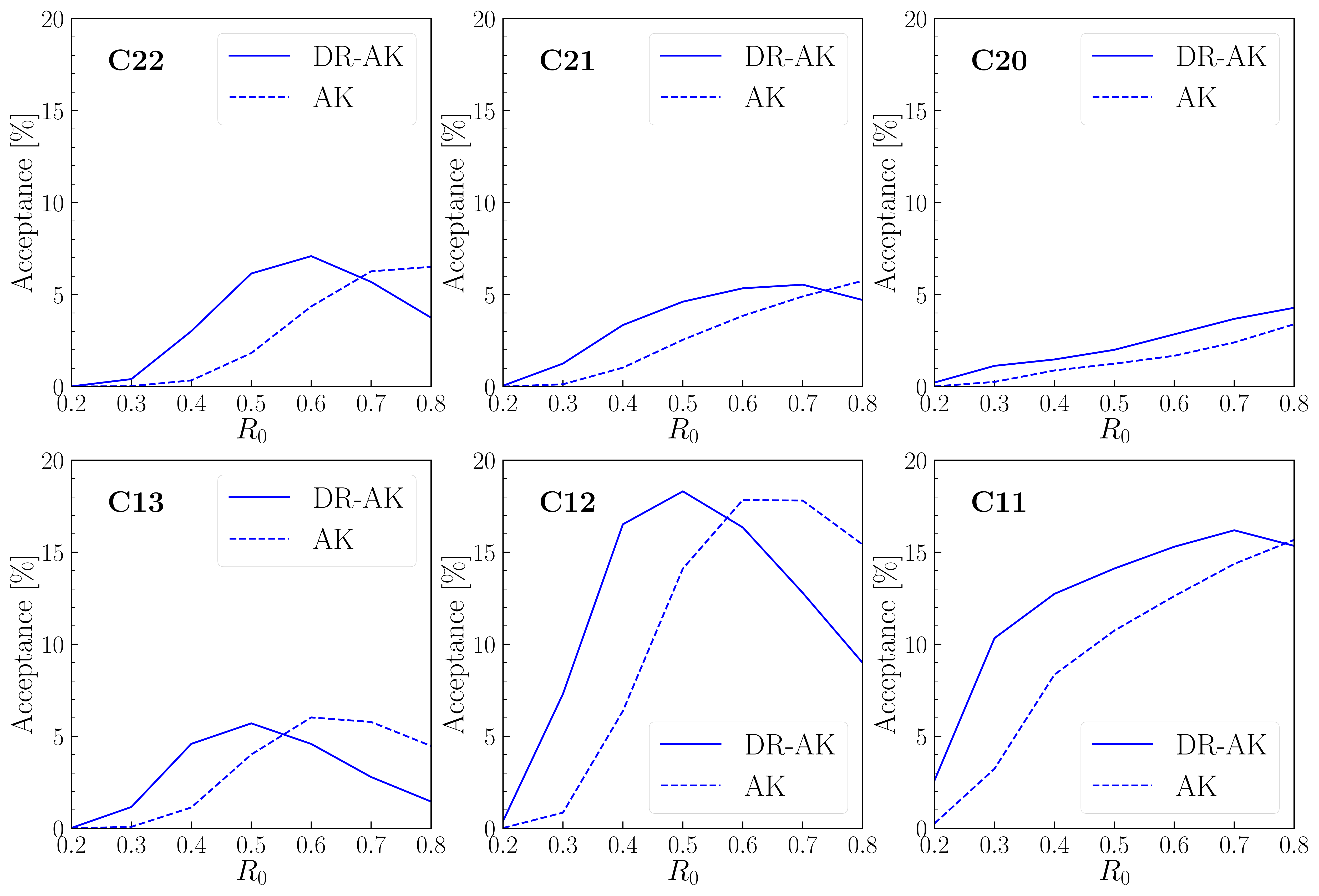}
\caption{The variation of \Acc [Eq.~(\ref{eqn:acc})] as a function of initial
radius $\Rnot$ for six categories, namely C22, C21, C20, C13, C12, and C11. The
solid blue lines are for the DR-AK algorithm, and the dashed lines are for the
AK algorithm. The jets are clustered with $\Rnot=0.5$.}
\label{fig:twtw-summary}
\end{center}\vspace{-5mm}
\end{figure}
%===============================================================================

The variation of \Acc as a function of initial radius $\Rnot$ is shown in
Fig.~\ref{fig:twtw-summary} for six categories, namely C22, C21, C20, C13, C12,
and C11. These categories have at least one \Top\ jet identified within 0.5
distance from the \MGpartonlevel top quark. The solid blue lines represent the
efficiencies for the DR-AK algorithm, and the dashed lines are representative of
the AK algorithm.  For the case of dynamic radius, the quintessential
feature is the initial increment in the acceptance efficiencies \Acc up to
$\Rnot=0.5$, and, beyond this value, the efficiencies decrease. The reason for
this is an unnecessary accumulation of hadrons and making the jets bigger than
their optimal size. However, for the AK algorithm, the efficiencies keep on
increasing until $\Rnot=0.7$, which is kind of the optimal radius for this
scenario. The most important point to note is that up to $\Rnot$=0.5, the
efficiencies for the DR-AK algorithm are higher than those for the AK algorithm.
This feature, again, establishes the utility of using dynamic radius algorithms
over fixed radius ones.

\begin{figure}[!h]
\begin{center}
\includegraphics[width=0.7\textwidth]{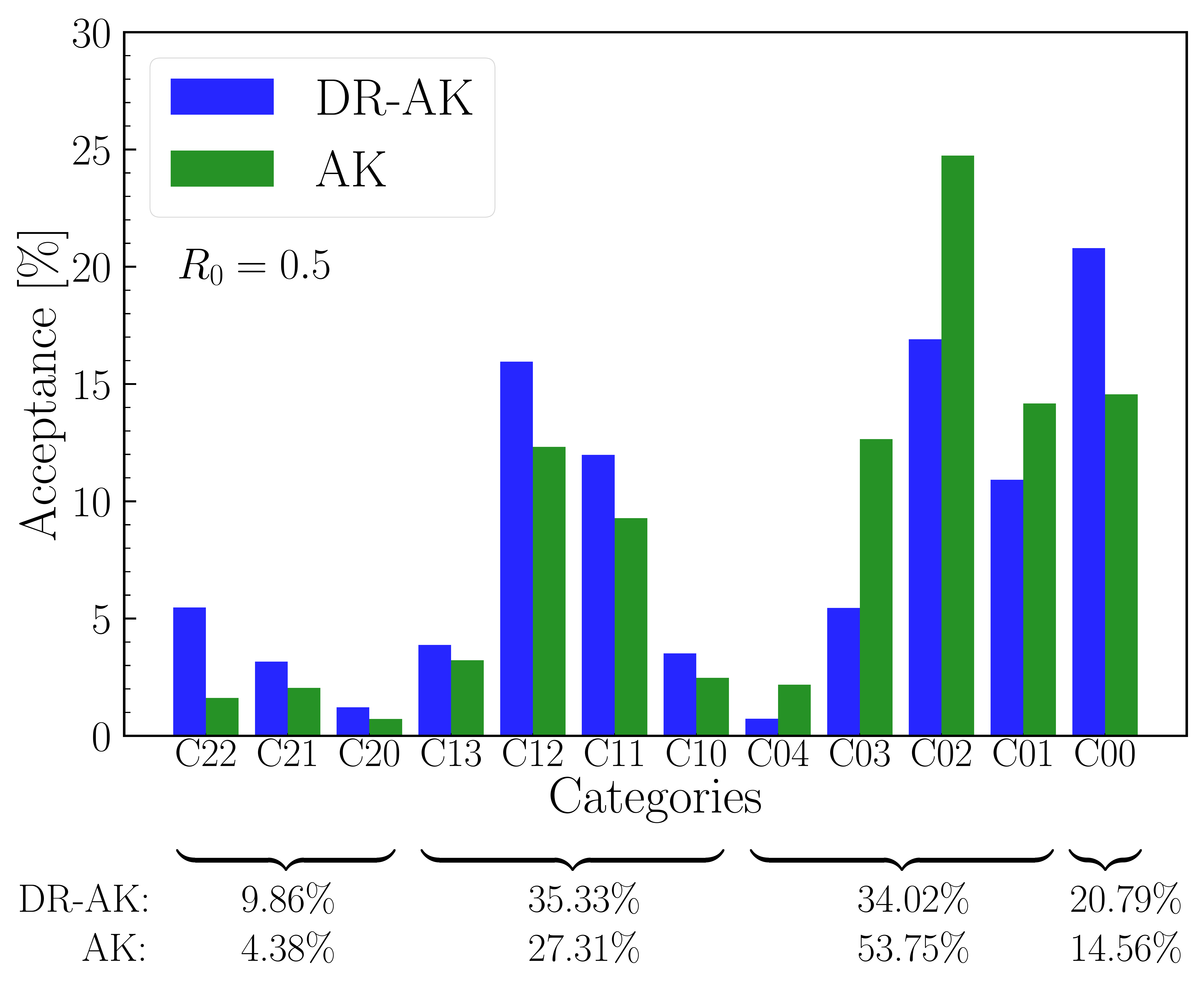}
\caption{Bar plot of \Acc for different categories for jet algorithms with
$\Rnot=0.5$. The blue, and green bars are for DR-AK and AK algorithm
respectively. From left to right, The numbers under the braces represent the
values of \Acc for the categories which capture 2 \Top, 1 \Top, 0 \Top, and
none of the \Top\ or \W\ jets.}
\label{fig:twtw-summary2}
\end{center}
\end{figure}

In the end, we look at the bar plot of the acceptance efficiencies \Acc for all
the categories in Fig.~\ref{fig:twtw-summary2}. The blue and green bars are for
DR-AK and AK algorithms, respectively. The initial radius $\Rnot$ is taken to
be $0.5$. The numbers under the curly braces below the $x$-axis represent the
values of \Acc for the categories which capture 2 \Top s, 1 \Top, 0 \Top, and
none of the \Top\ or \W\ jets. The important observation in this regard is that
the categories containing 2 \Top\ and 1 \Top\ jets have better efficiencies for
the dynamic radius algorithm than the fixed radius one. This means that the
events, where the AK algorithm could not capture the whole of the \Top\
constituents, the DR-AK algorithm could capture the full \Tops. Thus the
credence of our proposed algorithm is established in a BSM context as well. 

%===============================================================================
%===============================================================================

%===============================================================================
\section{Summary and Outlook}\label{sec:summary}
We go beyond the most popular jet clustering algorithms, where the formation of
jets is performed using a fixed radius parameter. These algorithms return
fixed-sized jets corresponding to the input radius parameter. In this work, an
attempt is made to make the radius of each jet variable depending on the
kinematics and hadronic activity in the neighbourhood of an evolving jet. The
proposed method is based on the standard $\kt$-type sequential recombination
jet clustering algorithms with the incorporation of the dynamic nature of the
radius parameter.

Starting from a reasonable radius parameter, during the process of formation of
a jet, the radius of each evolving jet is allowed to grow based on {\it
fuzziness} inside it. For this work, the measure of the fuzziness of each
evolving jet is chosen to be the `$\pt$-weighted' standard deviation of the
inter-particle distances (in the $\eta$-$\phi$ plane) of the particles inside
the evolving jet.

After describing the proposed method, we have presented two different SM
processes, {\it viz.} $pp\to tj$ and $pp\to Wj+Zj$, to demonstrate some
applicabilities of the dynamic radius jet clustering algorithm. In these two
processes, differently-sized jets are expected in a single event. In the two SM
process examples, we observe that the jets are being formed with radii varying
in size on a jet-by-jet basis. In terms of the acceptance efficiency
[Eq.~(\ref{eqn:acc})], we show that the performance of the dynamic radius
algorithm is better compared to their fixed radius counterparts. We take up a
scenario beyond the Standard Model for further illustration, where a vectorlike
$SU(2)_L$ singlet charge $-1/3$ quark $b'$ is added. We study jet clustering in
$pp\to b' \bar b'$ followed by each $b'$ decaying into $tW$. Once more, our
proposed method turns out to be effective in the reconstruction of the final
state particles. 

In the examples given above, the dynamicity has been incorporated in the radius
parameter of the standard $\kt$-type sequential recombination algorithm. The
central idea is the usage of {\it fuzziness} of an evolving jet to appropriately
increase its radius starting from a starting radius $\Rnot$. Although examples
with only one measure of {\it fuzziness} have been shown in this work, one may
consider other appropriate measures. depending upon the underlying physics
process or the final goal of the analysis. Therefore, the idea of the dynamic
radius jet algorithm should not be restricted only to this particular measure.
The applicability of these possibilities will be presented in a separate work.
In a nutshell, the idea of dynamic radius jet clustering algorithm on a
jet-by-jet basis is useful in collider studies and will be beneficial in
searches driven by processes in SM as well as BSM.

%===============================================================================
\acknowledgments
The authors thank Jayita Lahiri for useful discussions during the initial phase
of the work.

%===============================================================================
%\bibliographystyle{JHEP}
%\bibliography{ref.bib}
\input{ref.bbl}
\end{document}

%% file: ref.bbl
\providecommand{\href}[2]{#2}\begingroup\raggedright\endgroup